\documentclass[amssymb,twocolumn,aps]{revtex4-2}
\usepackage{graphicx}
\usepackage{enumerate}
\usepackage{amssymb}
\usepackage{amsmath}
\usepackage{amsfonts}
\usepackage{color}
\usepackage{mathrsfs}

\begin{document}

\title{Quantum electrodynamics of lossy magnetodielectric samples in vacuum: modified Langevin noise formalism}

\author{A. Ciattoni$^1$}
\email{alessandro.ciattoni@spin.cnr.it}
\affiliation{$^1$CNR-SPIN, c/o Dip.to di Scienze Fisiche e Chimiche, Via Vetoio, 67100 Coppito (L'Aquila), Italy}

\date{\today}

\begin{abstract}
Quantum behavior of the electromagnetic field in unbounded macroscopic media displaying absorption is properly described by the Langevin noise formalism (macroscopic quantum electrodynamics) where the field is assumed to be entirely produced by medium fluctuating sources via the dyadic Green's function. On the other hand, such formalism is able to deal with the case of finite-size lossy objects placed in vacuum only as a limiting situation where the permittivity limit ${\rm Im} ( \varepsilon) \rightarrow 0^+$ pertaining the regions filled by vacuum is taken at the end of the calculations. Strictly setting ${\rm Im} ( \varepsilon) =0$ is forbidden in the Langevin noise formalism since the field would vanish in the lossless regions and this is physically due to the fact that the contribution of the scattering modes to the field is not separated from the contribution produced by the medium fluctuating sources. Recently, a modified Langevin noise formalism has been proposed to encompass the scattering modes and accordingly it is able to describe the structured lossless situations by strictly setting ${\rm Im} (\varepsilon) = 0$. However such modified formalism has been numerically validated only in few specific geometries. In this paper we analytically derive the modified Langevin noise formalism from the established canonical quantization of the electromagnetic field in macroscopic media, thus proving that it models any possible scenario involving linear, inhomegeneous and magnetodielectric samples. The derivation starts from quantum Maxwell equations in the Heisenberg picture together with their formal solution as the superposition of the medium assisted field and the scattering modes. We analytically prove that each of the two field parts can be expressed in term of particular bosonic operators, which in turn diagonalize the electromagnetic Hamiltonian and whose associated quasi-particles are medium assisted and scattering polaritons, respectively. The key ingredient underpinning our reasoning is a peculiar integral relation linking the far field amplitude of the dyadic Green's function and the scattering modes, relation we rigorously derive and physically explain by identifying the scattering modes as fields generated by infinitely far dipole point sources. 
\end{abstract}
\maketitle

\section{Introduction}
Investigating the quantum features of light interacting with macroscopic matter \cite{Wester} is both conceptually significant and crucial for describing a plethora of relevant applications. In view of the practical impossibility to achieve a detailed miscroscopic description, the quantization of the electromagnetic field, when it experiences a continuum dielectric only through its permittivity and permeability, calls for a suitable model effectively accounting for matter dispersion and absorption. Quantum description of damping generally relies on the model inclusion of a bath with infinitely many degrees of freedom to dissipate energy \cite{Philb0} and such strategy is adopted in the Huttner–Barnett approach \cite{Huttn1,Huttn2}, in turn based of the Hopfield \cite{Hopfie} and Fano \cite{Fanooo} models, where matter polarization is represented by an harmonic oscillator coupled to a continuum of reservoir oscillator fields. Even though generalizations of the Huttner-Barnett approach have been proposed \cite{Sutto1,Sutto2,Kheir1}, specializing the method to an arbitrary electromagnetic geometry is still a challenging task since dielectric permittivies with different dispersive properties are usually in order. In this respect, various and more flexible quatization schemes have been proposed not including the matter oscillator but encoding an arbitrary medium electromagnetic response into the field-reservoir coupling term \cite{Matloo,Kheir2,Bhattt,Amoos1,Sutto3,Amoos2}. In addition to such approaches where quantization is canonically performed, i.e. commutation realtions are imposed to classical canonically conjugated variables, a phenomenological method \cite{Grune1,Schee1,Dungg1,Schee2,Buhma1} has been proposed which is based on the flucuation-dissipation theorem and which is usually referred to as macroscopic quantum electrodynamics (MQED) or Langevin noise formalism (LNF). In this approach the field is assumed to be generated by medium fluctuating sources, using the classical dyadic Green's function \cite{CheTai,Chewww}, and quantization is performed by imposing the standard vacuum quantum electrodynamics commutation relations ruling the electric and induction magnetic fields. This leads to the introduction of bosonic polariton operators, whereas the harmonic oscillator-like Hamiltonian is postulated to produce the correct time evolution of field operators in the Heisenberg picture. Major virtues of such approach are its relative semplicity and extreme flexibility resulting from the central role played by the Green's function through which the medium properties solely show up and whose detailed knowledge is generally only required at the eventual numerical evaluation stage. Due to its large versatility, LNF has been extensively exploited to describe a number of applications in many research subjects encompassing dispersion forces \cite{Buhma2,Philb1,Buhma3,Buhma4,Butch1,Fuchss,Fiedle}, quantum emitters decay \cite{Schee3,Dungg2,Dzsot1,Alpegg,Rivera,Hemmer,Wangg1,Wangg2,Kosikk}, cavity QED \cite{Khanb1,Khanb2,Dzsot2}, quantum nanophotonics \cite{Maroci,Hakam1,Hakam2,Kurman,Feistt} and fast electrons scattering \cite{DiGiu1,Hayun1,Kfirr1,Ciatt1,Ciatt2}. It is here essential to point out that, even though LNF has been introduced in a phenomenological and non-canonical way, Philbin has  provided a canonical approach for quantizing the field-reservoir system fully reproducing, and hence theoretically substantiating, the LNF \cite{Philbi}. 

The main LNF prescription consists in relating the field operators to the polaritonic operators through the Green's function, this physically amounting to processes where the field is generated by localized sources. Such prescription is fully adequate to deal with electrodynamics in unbounded absorbing media where any field exponentially vanishes far from the sources and the energy flow is forbidden at infinity so that radiation from localized sources is the only field generation mechanism. On the other hand, if a finite-size dielectric object in placed in vacuum, the electromagnetic energy carried by the radiation generated by localized source can reach indefinitely far regions. Conversely, it is possible to place the radiation source as far as desired from the object, this resulting in a second field excitation mechanism where radiation is ideally sent from infinity as plane waves which are subsequently scattered by the object. Such reasonining agrees with the early observations of Di Stefano \cite{DiStef}, lately emphasized by Drezet in Ref.\cite{Dreze1} and further analyzed in \cite{Dreze2} where a closer analysis of the fundamental integral relation involving the dyadic Green's function has unveiled an additional surface term (also discussed in Ref.\cite{Franke}) which, for a finite-size dielectric object placed in vacuum, persists even when the integration surface is brought to infinity. The observation has triggered renewed interest on the subject \cite{Dorie1,Dorie2,Fores1,Fores2,Naaaa1} and a modified Langevin noise formalism (MLNF) explicitly accounting for the scattering modes has been proposed and numerically checked in few specific geometries \cite{Naaaa2}. One of the main differences between the MLNF and the LNF is that the former enables to treat the situations comprising lossless spatial regions by strictly setting ${\rm Im} (\varepsilon) = 0$ in such regions whereas the latter deals with the lossless case only by taking the singular limit ${\rm Im} ( \varepsilon) \rightarrow 0^+$ at the end of the calculations, as shown in Ref.\cite{Hanson}.

In this paper we extensively analyze quantum electrodynamics of macroscopic finite-size absorbing objects in vacuum and we analitically validate the modified Langevin noise formalism by deriving it from the Philbin approach of Ref.\cite{Philbi} where canonical quantization of macroscopic electromagnetism is achieved. Specifically, instead of diagonalizing the electromagnetic Hamiltonian through the Fano method, we start from quantum Maxwell equations (resulting as Heisenberg equations for the field operators) and we argue that their most general formal solution is the superposition of the medium assisted field (particular solution, containing the Green's function) and the scattering modes (general solution of the source-free equations). As a first crucial result we prove that each field contribution can be expressed in terms of specific creation and annihilation operators, and this is done by assuming bosonic commutation relations for them and showing that all the field commutation relations are self-consistently reproduced. Such result stems from the exact cancellation, when  evaluating the field commutation relations, between the above discussed surface term (containing the dyadic Green's function) and a contribution produced by the scattering modes, a rigorous and highly non-trivial result in view of the different character of the two fields they stem from. We physically interpret such balance by identifying the scattering modes as fields genererated by dipole point sources
infinitely far from the object. Besides, using the field expressions, we show that the electromagnetic Hamiltonian is the sum of two harmonic oscillator-like Hamiltonians, as expected from Ref.\cite{Naaaa2}, pertaining two different kids of quasi-particles, medium assisted and scattering polaritons, respectively.

\section{Quantum Maxwell equations}
We start by reviewing the main results of the quantum theory of macroscopic electromagnetism discussed in Ref.\cite{Philbi}. The considered medium is an arbitrary isotropic, inhomogeneous   magnetodielectric one whose macroscopic  electromagnetic response is described by the polarization and magnetization densities
\begin{eqnarray} \label{PeMm}
 {\bf{P}}^{(\varepsilon)}  \left( {{\bf{r}},t} \right) &=& \frac{{\varepsilon _0 }}{{2\pi }}\int\limits_{ - \infty }^t {dt'} \;\chi ^{(\varepsilon)}  \left( {{\bf{r}},t - t'} \right){\bf{E}}\left( {{\bf{r}},t'} \right), \nonumber \\ 
 {\bf{M}}^{(\mu)}  \left( {{\bf{r}},t} \right) &=& \frac{1}{{2\pi \mu _0 }}\int\limits_{ - \infty }^t {dt'} \;\chi ^{(\mu)}  \left( {{\bf{r}},t - t'} \right){\bf{B}}\left( {{\bf{r}},t'} \right), 
\end{eqnarray}
where $\chi^{(\varepsilon)}({\bf r},\tau)$ and $\chi^{(\mu)} ({\bf r},\tau)$ are the electric and magnetic susceptibilities, respectively, whose vanishing for $\tau <0$ due to causality requires their frequency domain representations
\begin{equation}
\chi _\omega ^{(\varepsilon ,\mu )}({\bf r})  = \frac{1}{{2\pi }}\int\limits_0^\infty  {d\tau e^{i\omega \tau } \chi ^{(\varepsilon ,\mu )} \left( {{\bf{r}},\tau } \right)}
\end{equation}
to satisfy the Kramers-Kronig relations
\begin{eqnarray} \label{KK}
{\mathop{\rm Re}\nolimits} \, \chi _\omega ^{(\varepsilon ,\mu) }  = \frac{2}{\pi }P\int\limits_0^{ + \infty } {d\omega ' \, \frac{{\omega ' \, {\mathop{\rm Im}\nolimits} \, \chi _{\omega '}^{(\varepsilon ,\mu )} }}{{\omega '^2  - \omega ^2 }}}.
\end{eqnarray}
The electric permittivity and mangnetic permeability are defined in the frequency domain as
\begin{eqnarray}
 \varepsilon _\omega  \left( {\bf{r}} \right) &=& 1 + \chi _\omega ^{(\varepsilon)}  \left( {\bf{r}} \right), \nonumber \\ 
 \mu _\omega  \left( {\bf{r}} \right) &=& \frac{1} {1 - \chi _\omega ^{(\mu)}  \left( {\bf{r}} \right)},
 \end{eqnarray}
respectively.

The Heisenberg picture is used hereafter and 
 the electromagnetic field is described by the vector potential operator ${\bf{\hat A}}\left( {{\bf{r}},t} \right)$ in the Coulomb gauge $\nabla \cdot {\bf{\hat A}} = 0$. Besides, the reservoir is modelled  by the field operators ${\bf{\hat X}}^\Omega  \left( {{\bf{r}},t} \right)$ and ${\bf{\hat Y}}^\Omega  \left( {{\bf{r}},t} \right)$ representing a twofold continuum of harmonic oscillators of proper frequencies $\Omega$ and enabling the description of electric and magnetic energy dissipation, respectively. Letting $ {\bf{\hat \Pi }}_A \left( {{\bf{r}},t} \right)$, ${\bf{\hat \Pi }}_X^\Omega  \left( {{\bf{r}},t} \right)$ and ${\bf{\hat \Pi }}_Y^\Omega  \left( {{\bf{r}},t} \right)$ be the corresponding canonical momenta, the basic equal-time canonical commutation relations are 
\begin{eqnarray} \label{CanComRel}
 \left[ {{\bf{\hat A}}\left( {{\bf{r}},t} \right),{\bf{\hat \Pi }}_{A} \left( {{\bf{r}}',t} \right)} \right] &=& i\hbar \delta ^ \bot  \left( {{\bf{r}} - {\bf{r}}'} \right), \nonumber \\ 
 \left[ {{\bf{\hat X}}^\Omega  \left( {{\bf{r}},t} \right),{\bf{\hat \Pi }}_X^{\Omega '} \left( {{\bf{r}}',t} \right)} \right] &=& i\hbar \delta \left( {\Omega  - \Omega '} \right)\delta \left( {{\bf{r}} - {\bf{r}}'} \right)I, \nonumber \\ 
 \left[ {{\bf{\hat Y}}^\Omega  \left( {{\bf{r}},t} \right),{\bf{\hat \Pi }}_Y^{\Omega '} \left( {{\bf{r}}',t} \right)} \right] &=& i\hbar \delta \left( {\Omega  - \Omega '} \right)\delta \left( {{\bf{r}} - {\bf{r}}'} \right)I,
\end{eqnarray}
together with the vanishing of any other equal-time commutation relation (here $\delta ^ \bot  \left( {\bf{r}} \right)$ is the dyadic transverse delta function and $I$ is the dyadic identity, see Appendix A). The electromagnetic field-reservoir coupled dynamics is generated by the Hamiltonian
\begin{eqnarray} \label{Ham1}
&& \hat H = \int {d^3 {\bf{r}}} \left\{ {\frac{1}{{\varepsilon _0 }}{\bf{\hat \Pi }}_A  \cdot \left( {\frac{1}{2}{\bf{\hat \Pi }}_A  + \int\limits_0^{ + \infty } {d\Omega } \;\alpha _\Omega  {\bf{\hat X}}^\Omega  } \right) + } \right. \nonumber \\ 
&&  + \frac{1}{{2\varepsilon _0 }}\left( {\int\limits_0^{ + \infty } {d\Omega } \;\alpha _\Omega  {\bf{\hat X}}^\Omega  } \right)^2  + \frac{1}{{2\mu _0 }}\left| {\nabla  \times {\bf{\hat A}}} \right|^2  + \nonumber \\ 
&&  - \left( {\int\limits_0^{ + \infty } {d\Omega } \;\beta _\Omega  {\bf{\hat Y}}^\Omega  } \right) \cdot \left( {\nabla  \times {\bf{\hat A}}} \right) + \nonumber \\ 
&& \left. { + \frac{1}{2}\int\limits_0^{ + \infty } {d\Omega } \left[ {\left( {{\bf{\hat \Pi }}_X^{\Omega 2}  + {\bf{\hat \Pi }}_Y^{\Omega 2} } \right) + \Omega ^2 \left( {{\bf{\hat X}}^{\Omega 2}  + {\bf{\hat Y}}^{\Omega 2} } \right)} \right]} \right\}, \nonumber \\ 
\end{eqnarray}
where 
\begin{eqnarray} \label{coupCoeff}
 \alpha _\omega  \left( {\bf{r}} \right) &=& \sqrt {\frac{{2\varepsilon _0 }}{\pi }\omega {\mathop{\rm Im}\nolimits} \left[ {\varepsilon _\omega  \left( {\bf{r}} \right)} \right]}, \nonumber  \\ 
 \beta _\omega  \left( {\bf{r}} \right) &=& \sqrt {  \frac{2}{{\pi \mu _0 }}\omega {\mathop{\rm Im}\nolimits} \left[  {\frac{-1}{{\mu _\omega  \left( {\bf{r}} \right)}}} \right]}, 
\end{eqnarray}
are fundamental real coupling coefficients fully accounting for the medium response. Besides, the electric and induction magnetic field operators are given by
\begin{eqnarray} \label{EB}
 {\bf{\hat E}} &=&  - \frac{{\partial {\bf{\hat A}}}}{{\partial t}} + \left( - \frac{1}{{\varepsilon _0 }} {\int\limits_0^{ + \infty } {d\Omega } \;\alpha _\Omega  {\bf{\hat X}}^\Omega  } \right)^\parallel, \nonumber \\ 
 {\bf{\hat B}} &=& \nabla  \times {\bf{\hat A}},
\end{eqnarray}
where the superscript $\parallel$ labels the longitudinal part of the vector (see Appendix A), so that the homogeneous quantum Maxwell equations
\begin{eqnarray} \label{homMax}
\nabla  \cdot {\bf{\hat B}} &=& 0 , \nonumber \\ 
\nabla  \times {\bf{\hat E}} &=&  - \frac{{\partial {\bf{\hat B}}}}{{\partial t}}
\end{eqnarray}
are automatically satisfied. Such quantum description arises from a canonical quantization procedure which, after exhibiting a suitable Lagragian whose Euler-Lagrange equations reproduce classical macroscopic Maxwell equations, indentifies conjugated variables to which canonical commutation relations are imposed (constrained by the transverse character of the vector potential in the Coulomb gauge). 

An additional and remarkable feature of the approach is that the Heisenberg equations of the canonical field operators yield the inhomogenous quantum Maxwell equations, as pointed out in Ref.\cite{Philbi}. Since this is the starting point of the present paper investigation, we closely review this result by explicitly writing  the six dynamical Heisenberg equations
\begin{eqnarray} \label{Heisen}
 \frac{{\partial {\bf{\hat A}}}}{{\partial t}} &=& \frac{1}{{\varepsilon _0 }}{\bf{\hat \Pi }}_A  + \left( {\frac{1}{{\varepsilon _0 }}\int\limits_0^{ + \infty } {d\Omega } \;\alpha _\Omega  {\bf{\hat X}}^\Omega  } \right)^ \bot , \nonumber   \\ 
\frac{{\partial {\bf{\hat \Pi }}_A }}{{\partial t}} &=& \;\nabla  \times \left( { - \frac{1}{{\mu _0 }}\nabla  \times {\bf{\hat A}} + \int\limits_0^{ + \infty } {d\Omega } \;\beta _\Omega  {\bf{\hat Y}}^\Omega  } \right) , \nonumber   \\ 
 \frac{{\partial {\bf{\hat X}}^\Omega  }}{{\partial t}} &=& {\bf{\hat \Pi }}_X^\Omega, \nonumber   \\ 
 \frac{{\partial {\bf{\hat \Pi }}_X^\Omega  }}{{\partial t}} &=&  - \frac{{\alpha _\Omega  }}{{\varepsilon _0 }}\left( {{\bf{\hat \Pi }}_{A}  + \int\limits_0^{ + \infty } {d\Omega '} \;\alpha _{\Omega '} {\bf{\hat X}}^{\Omega '} } \right) - \Omega ^2 {\bf{\hat X}}^\Omega, \nonumber   \\ 
 \frac{{\partial {\bf{\hat Y}}^\Omega  }}{{\partial t}} &=& {\bf{\hat \Pi }}_Y^\Omega, \nonumber   \\ 
 \frac{{\partial {\bf{\hat \Pi }}_Y^\Omega  }}{{\partial t}} &=&  - \Omega ^2 {\bf{\hat Y}}^\Omega   + \;\beta _\Omega  \nabla  \times {\bf{\hat A}}, 
\end{eqnarray}
where the superscript $\bot$ labels the transverse part of the vector (see Appendix A), and which are obtained by using the general equation $\frac{{\partial \hat O}}{{\partial t}} = \frac{i}{\hbar } [ {\hat H,\hat O}]$ for the Heisenberg operator $\hat O(t)$ together with the commutation relations of Eqs.(\ref{CanComRel}). Combining the first of Eqs.(\ref{EB}) and the first of Eqs.(\ref{Heisen}) we get 
\begin{equation} \label{PA}
{\bf{\hat \Pi }}_A  =  - \varepsilon _0 {\bf{\hat E}} - \int\limits_0^{ + \infty } {d\Omega } \;\alpha _\Omega  {\bf{\hat X}}^\Omega  
\end{equation}
which, again from the first of Eqs.(\ref{Heisen}), is easily seen to be a transverse field operator (in agreement with the first of Eqs.(\ref{CanComRel})), this leading to the equation
\begin{equation} \label{Max1}
\nabla  \cdot \left( {\varepsilon _0 {\bf{\hat E}} + \int\limits_0^{ + \infty } {d\Omega } \;\alpha _\Omega  {\bf{\hat X}}^\Omega  } \right) = 0.
\end{equation}
Besides, using the second of Eqs.(\ref{EB}) together with Eq.(\ref{PA}), the second of Eqs.(\ref{Heisen}) turns into
\begin{eqnarray} \label{Max2}
&& \nabla  \times \left( { - \frac{1}{{\mu _0 }}{\bf{\hat B}} + \int\limits_0^{ + \infty } {d\Omega } \;\beta _\omega  {\bf{\hat Y}}^\Omega  } \right) + \nonumber \\
&& +\frac{\partial }{{\partial t}}\left( {\varepsilon _0 {\bf{\hat E}} + \int\limits_0^{ + \infty } {d\Omega } \;\alpha _\Omega  {\bf{\hat X}}^\Omega  } \right) = 0.
\end{eqnarray}
Equations (\ref{Max1}) and (\ref{Max2}) are seen to be the inhomogeneous quantum Maxwell equations. To prove this, eliminating  reservoir fields is essential and this can be done by noting that, after using Eq.(\ref{PA}) together with the second of Eqs.(\ref{EB}), the last four of Eqs.(\ref{Heisen}) yield the forced harmonic oscillator equations
\begin{eqnarray} \label{XYOscil}
 \frac{{\partial ^2 {\bf{\hat X}}^\Omega  }}{{\partial t^2 }} + \Omega ^2 {\bf{\hat X}}^\Omega   &=& \alpha _\Omega  {\bf{\hat E}}, \nonumber \\ 
 \frac{{\partial ^2 {\bf{\hat Y}}^\Omega  }}{{\partial t^2 }} + \Omega ^2 {\bf{\hat Y}}^\Omega   &=& \;\beta _\Omega  {\bf{\hat B}},
\end{eqnarray}
whose most general and causal solutions are
\begin{eqnarray} \label{XOYO}
&& {\bf{\hat X}}^\Omega  \left( {{\bf{r}},t} \right) = \alpha _\Omega  \left( {\bf{r}} \right)\int\limits_{ - \infty }^t {dt'} \,\,\frac{{\sin \left[ {\Omega \left( {t - t'} \right)} \right]}}{\Omega }{\bf{\hat E}}\left( {{\bf{r}},t'} \right) + \nonumber \\ 
&& + \left[ {e^{ - i\Omega t} {\bf{\hat Z}}_\Omega  \left( {\bf{r}} \right) + {\rm h.c.}} \right], \nonumber \\ 
&& {\bf{\hat Y}}^\Omega  \left( {{\bf{r}},t} \right) = \beta _\Omega  \left( {\bf{r}} \right)\int\limits_{ - \infty }^t {dt'} \,\,\frac{{\sin \left[ {\Omega \left( {t - t'} \right)} \right]}}{\Omega }{\bf{\hat B}}\left( {{\bf{r}},t'} \right) + \nonumber \\ 
&& +  \left[ {e^{ - i\Omega t} {\bf{\hat W}}_\Omega  \left( {\bf{r}} \right) + {\rm h.c.}} \right].
\end{eqnarray}
Here the terms containing the time integrals provide the particular solutions by means of the retarded Green's function of the classical harmonic oscillator whereas the remaining terms containing the operators ${\bf{\hat Z}}_\Omega$ and ${\bf{\hat W}}_\Omega$ represent the general solutions of the homogeneous equations. Now, using the polarization and magnetization densities of Eqs.(\ref{PeMm}), the Kramers-Kronig relation of Eq.(\ref{KK}) and the frequency domain representation of the operators 
\begin{equation} \label{HeisenOpe}
{\bf{\hat F}}\left( {{\bf{r}},t} \right) = \int\limits_0^{ + \infty } {d\omega e^{ - i\omega t} } {\bf{\hat F}}_\omega  \left( {\bf{r}} \right) + {\rm h.c.},
\end{equation}
it is possible to prove the relations
\begin{eqnarray} \label{intFre}
 \int\limits_0^{ + \infty } {d\Omega } \;\alpha _\Omega  {\bf{\hat X}}^\Omega   = {\bf{\hat P}}^{(\varepsilon)}   + {\bf{\hat P}}^{(M)}, \nonumber  \\ 
 \int\limits_0^{ + \infty } {d\Omega } \;\beta _\Omega  {\bf{\hat Y}}^\Omega   = {\bf{\hat M}}^{(\mu)}   + {\bf{\hat M}}^{(M)} , 
\end{eqnarray}
where
\begin{eqnarray} \label{PNMN}
{\bf{\hat P}}^{(M)}_{\omega } = \alpha _\omega  {\bf{\hat Z}}_\omega, \nonumber\\
{\bf{\hat M}}^{(M)}_{\omega }  = \beta _\omega  {\bf{\hat W}}_\omega, 
\end{eqnarray}
are the noise polarization and magnetization densities in the frequency domain, respectively. Equations (\ref{intFre}) clarify the physical meaning of the reservoir fields by showing that $\alpha_\Omega {\bf{\hat X}}^\Omega$ and $\beta_\Omega {\bf{\hat Y}}^\Omega$ are the differential contributions to the total polarization and magnetization densities of the medium, comprising the permittivity and permeability parts ${\bf{\hat P}}^{(\varepsilon)}$,${\bf{\hat M}}^{(\mu)}  $ and the fluctuating noise source parts ${\bf{\hat P}}^{(M)}$,${\bf{\hat M}}^{(M)}$. We hereafter label with the superscript $(M)$ any physical quantity which is related with such medium assisted flucuating sources. Inserting Eqs.(\ref{intFre}) into Eqs.(\ref{Max1}) and (\ref{Max2}) we eventually get the inhomogeneous quantum Maxwell equations
\begin{eqnarray} \label{QMeq}
 \nabla  \cdot {\bf{\hat D}} &=& \hat \rho ^{(M)}, \nonumber  \\ 
 \nabla  \times {\bf{\hat H}} &=& {\bf{\hat J}}^{(M)}  + \frac{{\partial {\bf{\hat D}}}}{{\partial t}},
\end{eqnarray}
where 
\begin{eqnarray} \label{DH}
{\bf{\hat D}} &=& \varepsilon _0 {\bf{\hat E}} + {\bf{\hat P}}^{(\varepsilon)}   , \nonumber \\
{\bf{\hat H}} &=& \frac{1}{{\mu _0 }}{\bf{\hat B}} - {\bf{\hat M}}^{(\mu)}
\end{eqnarray} 
are the electric displacement and magnetic field , respectively, whereas 
\begin{eqnarray} \label{NoiSou}
 \hat \rho ^{(M)}  &=&  - \nabla  \cdot {\bf{\hat P}}^{(M)}, \nonumber  \\ 
 {\bf{\hat J}}^{(M)}  &=& \frac{{\partial {\bf{\hat P}}^{(M)} }}{{\partial t}} + \nabla  \times {\bf{\hat M}}^{(M)}
\end{eqnarray}
are the noise charge and current densities 
evidently satisfying the continuity equation $\nabla  \cdot {\bf{\hat J}}^{(M)}  + \frac{{\partial \hat \rho ^{(M)} }}{{\partial t}} = 0$. It is worth stressing that the above quantum Maxwell equations are fully general and they hold in every medium geometry also possibly encompassing spatial regions filled by vacuum where ${\bf{\hat D}} = \varepsilon _0 {\bf{\hat E}}$, ${\bf{\hat H}} = \frac{1}{{\mu _0 }}{\bf{\hat B}}$ and the noise sources of Eqs.(\ref{NoiSou}) vanish due to the coupling coefficients $\alpha_\omega$ and $\beta_\omega$ in the expressions for ${\bf{\hat P}}^{(M)}_{\omega }$ and ${\bf{\hat M}}^{(M)}_{\omega }$ in Eqs.(\ref{PNMN}).

As a final but not less important comment, we remark the central role played by the operators  
${\bf{\hat Z}}_\Omega$ and ${\bf{\hat W}}_\Omega$
which appeared in the Eqs.(\ref{XOYO}) merely to represent the possible free evolution part of the reservoir fields ${\bf{\hat X}}^\Omega$ and ${\bf{\hat Y}}^\Omega $. After obtaining quantum Maxwell equations of Eqs.(\ref{QMeq}), the operators ${\bf{\hat Z}}_\Omega$ and ${\bf{\hat W}}_\Omega$ have been found to be associated with additional charge $\hat \rho ^{(M)}$ and current $ {\bf{\hat J}}^{(M)}$ densities and it is remarkable that, in classical optics, their classical counterparts are usually dramatically ignored. Conversely, in the quantum approach we are considering, such seemingly superfluous quantum sources can not be arbitrarily suppressed without violating the canonical commutation relations (see the end of Sec.IV) and are essential to describe the part of the field which is produced by localized sources, the medium assisted field (see below).

We conclude this paragraph by reporting the frequency domain representation of the field operators and of their conjugated canonical momenta 
\begin{eqnarray} \label{FreqRepFiel}
 {\bf{\hat A}}_\omega   &=& \frac{1}{{i\omega }}{\bf{\hat E}}_\omega ^ \bot , \nonumber \\ 
 {\bf{\hat X}}_\omega ^\Omega   &=& \frac{{\alpha _\Omega  }}{{\Omega ^2  - \left( {\omega  + i\eta } \right)^2 }}{\bf{\hat E}}_\omega   + \delta \left( {\omega  - \Omega } \right){\bf{\hat Z}}_\omega  , \nonumber \\ 
 {\bf{\hat Y}}_\omega ^\Omega   &=& \frac{{\beta _\Omega  }}{{\Omega ^2  - \left( {\omega  + i\eta } \right)^2 }}\frac{{\nabla  \times {\bf{\hat E}}_\omega  }}{{i\omega }} + \delta \left( {\omega  - \Omega } \right){\bf{\hat W}}_\omega , \nonumber   \\ 
 \left( {{\bf{\hat \Pi }}_{A} } \right)_\omega   &=&  - \varepsilon _0 \varepsilon _\omega  {\bf{\hat E}}_\omega   - \alpha _\omega  {\bf{\hat Z}}_\omega , \nonumber  \\ 
 \left( {{\bf{\hat \Pi }}_{X}^\Omega } \right)_\omega   &=&  - i\omega {\bf{\hat X}}_\omega ^\Omega , \nonumber  \\ 
 \left( {{\bf{\hat \Pi }}_{Y}^\Omega } \right)_\omega   &=&  - i\omega {\bf{\hat Y}}_\omega ^\Omega  
\end{eqnarray}
where the prescription $\eta \rightarrow 0^+$ here accounts for the causality of the electromagnetic response.

\section{Medium assisted and scattering fields}
We now start our investigation which here departs from the approach described in Ref.\cite{Philbi} in that we directly analyze quantum Maxwell equations and their solutions instead of using the Fano method for diagonalizing the Hamiltonian of Eq.(\ref{Ham1}). Quantum Maxwell equations are formally identical to their classical counterpart and hence they are better dealt with in the frequency domain where, in the usual way, they provide the inhomogeneous Helmholtz equation for the electric field
\begin{equation} \label{Helm}
\left( {\nabla  \times \frac{1}{{\mu _\omega  }}\nabla  \times  - {k_\omega^2}
 \varepsilon _\omega  } \right){\bf{\hat E}}_\omega   = i\omega \mu _0 {\bf{\hat J}}^{(M)}_{\omega },
\end{equation}
where $k_\omega = \omega /c$ is the vacuum wavenumber and 
\begin{equation} \label{JNo}
{\bf{\hat J}}_\omega ^{(M)}  =  - i\omega \alpha _\omega  {\bf{\hat Z}}_\omega   + \nabla  \times \left( {\beta _\omega  {\bf{\hat W}}_\omega  } \right).
\end{equation}
The general solution of Eq.(\ref{Helm}) is necessarily given by 
\begin{equation} \label{EENES}
{\bf{\hat E}}_\omega   = {\bf{\hat E}}_\omega ^{(M)}  + {\bf{\hat E}}_\omega ^{(S)}
\end{equation}
where ${\bf{\hat E}}_\omega ^{(M)}$ is the particular solution accounting for the noise source $i\omega \mu _0 {\bf{\hat J}}_\omega ^{(M)}$ whereas ${\bf{\hat E}}_\omega ^{(S)}$ is the most general solution of the homogeneous Helmholtz equation. Physically ${\bf{\hat E}}_\omega ^{(M)}$ represent the medium assisted field, i.e. the part of the field generated by the medium fluctuating noise sources whereas ${\bf{\hat E}}_\omega ^{(S)}$ is the part of the field which is produced by infinitely far sources so that it necessarily has to represent any possible scattering process. 
Now in the standard LNF only the medium assisted contribution ${\bf{\hat E}}_\omega ^{(M)}$ is phenomenologically assumed \cite{Grune1,Schee1,Dungg1,Schee2,Buhma1} or theoretically obtained by the Fano diagonalization method \cite{Philbi}. On the other hand, superposing the particular solution and the general solution of the homogeneous equation is a fundamental rule associated with the linearity of the equation and hence it is also rigorously true for operator equations as Eq.(\ref{Helm}) (or Eqs.(\ref{XYOscil}) as we have reviewed above). Therefore the introduction of the scattering field ${\bf{\hat E}}_\omega ^{(S)}$ in the MLNF is here not postulated but it is a direct consequence of quantum Maxwell equations. It is worth highlighting that, in the situation where the lossy medium fills the whole space, the scattering contribution ${\bf{\hat E}}_\omega ^{(S)}$ necessarily vanishes since solutions of the homogeneous Helmholtz equation exponentially growing at infinity are ruled out. From a physical point of view this amounts to the impossibility of finite amount of energy coming from infinity to reach finite portion of the lossy medium. In view of this observation, we hereafter specialze our investigation to the situation where an arbitrary magnetodielectric sample is placed in vacuum so that the coexistence of both medium assisted ${\bf{\hat E}}_\omega ^{(M)}$ and scattering ${\bf{\hat E}}_\omega ^{(S)}$ fields can be investigated. As a prelude to the quantum description of such coexistence, it is imperative to investigate both the properties of such two field contributions and their intimate relationship.

\subsection{Medium assisted field}
The medium assisted field is given by
\begin{equation} \label{ENo1}
{\bf{\hat E}}_\omega ^{(M)} \left( {\bf{r}} \right) = i\omega \mu _0 \int {d^3 {\bf{r}}'} {\cal G}_\omega  \left( {{\bf{r}},{\bf{r}}'} \right) \cdot {\bf{\hat J}}_\omega ^{(M)} \left( {{\bf{r}}'} \right)
\end{equation}
where ${\cal G}_\omega  \left( {{\bf{r}},{\bf{r}}'} \right)$ is the dyadic Green's function satisfying the equation
\begin{equation} \label{GreenEQ}
\left( {\nabla  \times \frac{1}{{\mu _\omega  }}\nabla  \times  - {k_\omega^2}
 \varepsilon _\omega  } \right){\cal G}_\omega  \left( {{\bf{r}},{\bf{r}}'} \right) = \delta \left( {{\bf{r}} - {\bf{r}}'} \right)I
\end{equation}
and the boundary condition ${\cal G}_\omega  \left( {{\bf{r}},{\bf{r}}'} \right) \to 0$ when $r,r' \to \infty$. In the specific situation we are considering in this paper where vacuum surrounds the sample, the dyadic Green's function asymptotically satisfies the Sommerfeld radiation condition for $r\to \infty $ espressed by the relations
\begin{eqnarray} \label{GreenAsym}
 {\cal G}_\omega  \left( {{\bf{r}},{\bf{r}}'} \right) &=& \frac{{e^{i k_\omega r} }}{r}{\cal W}_\omega  \left( {o_{\bf r} ,{\bf{r}}'} \right) + O\left( {\frac{1}{{r^2 }}} \right), \nonumber \\ 
 \nabla  \times {\cal G}_\omega  \left( {{\bf{r}},{\bf{r}}'} \right) &=& \frac{{e^{ik_\omega  r} }}{r}ik_\omega  {\bf{u}}_{\bf r}  \times {\cal W}_\omega  \left( {o_{\bf r} ,{\bf{r}}'} \right) + O\left( {\frac{1}{{r^2 }}} \right), \nonumber \\
\end{eqnarray}
where ${\bf{u}}_{\bf r}  = {\bf{r}}/r$ is the radial unit vector along the observation direction whose polar angles are $o_{\bf r}  = \left( {\theta _{\bf r} ,\varphi _{\bf r} } \right)$ and ${\cal W}_\omega  \left( {o_{\bf r} ,{\bf{r}}'} \right)$ is a dyadic amplitude independent on $r$ which is left-orthogonal to the observation direction, i.e. 
\begin{equation} \label{tranW}
{\bf{u}}_{\bf r}  \cdot {\cal W}_\omega  \left( {o_{\bf r} ,{\bf{r}}'} \right) = 0.
\end{equation}
The validity and self-consistency of the MLNF we are to prove in the present paper are almost entirely substantiated by two basic properties of the dyadic Green's function (which we prove in Appendix B), namely, the reciprocity relation
\begin{equation} \label{RecipG}
{\cal G}_\omega  \left( {{\bf{r}} ,{\bf{r}}' } \right) = {\cal G}_\omega ^T \left( {{\bf{r}}' ,{\bf{r}} } \right)
\end{equation}
and the fundamental integral relation
\begin{eqnarray} \label{FunIntG}
&& \int {d^3 {\bf{s}}} \sum\limits_{\lambda  = e,m} {{\cal A}_{\omega \lambda } \left( {{\bf{r}} ,{\bf{s}}} \right) \cdot {\cal A}_{\omega \lambda }^{*T} \left( {{\bf{r}}' ,{\bf{s}}} \right)}  + \nonumber  \\ 
&& + k_\omega  \int {do } \: {\cal W}_\omega ^T \left( {o ,{\bf{r}} } \right) \cdot {\cal W}_\omega ^* \left( {o ,{\bf{r}}' } \right) = \nonumber \\
&& = {\mathop{\rm Im}\nolimits} \left[ {{\cal G}_\omega  \left( {{\bf{r}} ,{\bf{r}}' } \right)} \right] 
\end{eqnarray}
where $o = \left( {\theta ,\varphi  } \right)$, $do = \sin \theta  d\theta  d\varphi $ is the solid angle differential, the integration is performed over the whole solid angle and we have introduced the dyadics
\begin{eqnarray} \label{AeAm}
 {\cal A}_{\omega e} \left( {{\bf{r}},{\bf{r}}'} \right) &=& k_\omega \sqrt {{\mathop{\rm Im}\nolimits} \left[ {\varepsilon _\omega  \left( {{\bf{r}}'} \right)} \right]} {\cal G}_\omega  \left( {{\bf{r}},{\bf{r}}'} \right), \nonumber \\ 
 {\cal A}_{\omega m} \left( {{\bf{r}},{\bf{r}}'} \right) &=&  \sqrt {{\mathop{\rm Im}\nolimits} \left[ {\frac{{ - 1}}{{\mu _\omega  \left( {{\bf{r}}'} \right)}}} \right]} \left[ {{\cal G}_\omega  \left( {{\bf{r}},{\bf{r}}'} \right) \times \mathord{\buildrel{\lower3pt\hbox{$\scriptscriptstyle\leftarrow$}} 
\over \nabla } _{{\bf{r}}'} } \right]. 
\end{eqnarray}
As correctly argumented by Drezet in Ref.\cite{Dreze2}, a surface contribution to the fundamental integral relation (second term in the left hand side of Eq.(\ref{FunIntG})) persists when the medium is not lossy and, here, we have extended the correctness of such basic observation to the more general magnetodielectric situation. 

In Appendix C we analytically check the fundamental integral relation of Eq.(\ref{FunIntG}) in the case where there is no lossy medium, i.e. the whole space is filled by vacuum. The result is particularly interesting for our purposes since for vacuum the dyadics of Eqs.(\ref{AeAm}) rigorously vanish so that the fundamental integral relation reduces to 
\begin{equation} \label{FunIntGvac}
k_\omega  \int {do} \;{\cal W}_\omega ^T \left( {o,{\bf{r}}} \right) \cdot {\cal W}_\omega ^* \left( {o,{\bf{r}}'} \right) = {\mathop{\rm Im}\nolimits} \left[ {{\cal G}_\omega  \left( {{\bf{r}},{\bf{r}}'} \right)} \right]
\end{equation}
i.e. its consistency is fully due to the surface contribution. Now this result, pertaining the MLNF we are describing here, should be compared with the analogous one derived from the standard LNF and discussed in Ref.\cite{Hanson} where the fundamental integral relation for the dyadic Green's function correctly lacks the surface contribution since there the medium is absorbing everywhere. In that analysis the lossless case is correctly obtained by taking the limit ${\rm Im} \left( \varepsilon_\omega \right) \rightarrow 0^+$ of the fundamental integral relation. We conclude that, as above stated, the lossless situation can be dealt with in the MLFN without retaining a small imaginary part of the permittivity (i.e. by setting ${\rm Im} \left( \varepsilon_\omega \right) = 0$ from the beginning) whereas in the LNF the limit ${\rm Im} \left( \varepsilon_\omega \right) \rightarrow 0^+$ has necessarily to be taken at the end of the calculations.

Inserting Eq.(\ref{JNo}) into Eq.(\ref{ENo1}) and using Eq.(\ref{DrotF}) of Appendix A together with the defintion of the coupling coefficients in Eqs.(\ref{coupCoeff}) we get
\begin{eqnarray} \label{ENo2}
&& {\bf{\hat E}}_\omega ^{(M)} \left( {\bf{r}} \right) = \sqrt {\frac{{2\mu _0 }}{\pi }\omega ^3 } \int {d^3 {\bf{r}}'}  \left[ {{\cal A}_{\omega e} \left( {{\bf{r}},{\bf{r}}'} \right) \cdot {\bf{\hat Z}}_\omega  \left( {{\bf{r}}'} \right) + } \right. \nonumber \\ 
&& + \left. {  i {\cal A}_{\omega m} \left( {{\bf{r}},{\bf{r}}'} \right) \cdot {\bf{\hat W}}_\omega  \left( {{\bf{r}}'} \right)} \right],
\end{eqnarray}
where we have dropped the surface term appearing in Eq.(\ref{DrotF}) since $\beta_\omega$ vanishes at infinity. Equation (\ref{ENo2}) directly relates the medium assisted field ${\bf{\hat E}}_\omega ^{(M)}$ to the operators ${{\bf{\hat Z}}_\omega  }$ and ${{\bf{\hat W}}_\omega  }$ pertaining the medium electric and magnetic noise sources, the link been provided by the electric and magnetic dyadics ${\cal A}_{\omega e}$ and ${\cal A}_{\omega m}$ which, due to the permittivity and permeability imaginary parts in Eqs.(\ref{AeAm}), vanish oustide the sample volume, thus effectively enabling the restriction of the integration domain in Eq.(\ref{ENo2}) to such volume.

\subsection{Scattering field}
The scattering field ${\bf{\hat E}}_\omega ^{(S)}$ is the most general solution of the homogeneous Helmholtz equation, namely
\begin{equation} \label{homHelm}
\left( {\nabla  \times \frac{1}{{\mu _\omega  }}\nabla  \times  - k_\omega^2 \varepsilon _\omega  } \right){\bf{\hat E}}_\omega ^{(S)}  = 0.
\end{equation}
An incident $(in)$ plane wave launched from vacuum onto the sample produces a scattered field $(sc)$ so that their superposition is a basic scattering mode $(S)$ which satisfies the homogeneous Helmholtz equation and it is labelled by the incident plane wave parameters. We express the incident plane wave as 
\begin{equation} \label{Fi}
{\bf{F}}_{\omega {\bf{n}} \nu }^{(in)} \left( {\bf{r}} \right) = e^{i k_\omega {\bf{n}} \cdot {\bf{r}}} {\bf{e}}_{{\bf{n}}\nu}, 
\end{equation}
where $\bf{n}$ is a unit vector along the wave vector ${\bf{k}} = k_\omega {\bf{n}}$ and ${\bf{e}}_{{\bf{n}}1} ,{\bf{e}}_{{\bf{n}}2}$ are two mutually orthogonal polarization unit vectors which are orthogonal to $\bf n$. Denoting with ${\bf{F}}_{\omega {\bf{n}}\nu }^{(sc)} \left( {\bf{r}} \right)$ the scattered field produced by the plane wave ${\bf{F}}_{\omega {\bf{n}}\nu }^{(in)}$, we here express the overall scattering mode as
\begin{equation} \label{FSKl}
{\bf{F}}_{\omega {\bf{n}}\nu }^{(S)} \left( {\bf{r}} \right) = {\bf{F}}_{\omega {\bf{n}}\nu }^{(in)} \left( {\bf{r}} \right) + {\bf{F}}_{\omega {\bf{n}}\nu }^{(sc)} \left( {\bf{r}} \right).
\end{equation}
Now the scattering field operator is constructed by superimposing all the possibile scattering modes, i.e. we set
\begin{equation} \label{ES}
{\bf{\hat E}}_\omega ^{(S)} \left( {\bf{r}} \right) = \int {do_{\bf n} } \sum\limits_{\nu =1,2}  {{\bf{F}}_{\omega {\bf{n}}\nu }^{(S)} \left( {\bf{r}} \right)\hat Q_{\omega {\bf{n}}\nu } } 
\end{equation}
where $o_{\bf n}  = \left( {\theta _{\bf n} ,\varphi _{\bf n} } \right)$ are the polar angles of the unit vector $\bf n$, $do_{\bf n}  = \sin \theta _{\bf n} d\theta _{\bf n} d\varphi _{\bf n}$ is the solid angle differential, the integration is performed over the whole solid angle and $\hat Q_{\omega {\bf{n}}\nu }$ are operators. 

Before moving on to the validation of the full MLNF, it is here essential discussing the scattering modes ${\bf{F}}_{\omega {\bf{n}}\nu }^{(S)}$  together with some of their remarkable properties. We start by noting that in the Eq.(\ref{FSKl}) we have defined the scattered field ${\bf{F}}_{\omega {\bf{n}}\nu }^{(sc)} $ in such a way that this relation holds throughout the whole space (both in vacuum and inside the sample). Now this is particularly convenient since, using the fact that the scattering mode field ${\bf{F}}_{\omega {\bf{n}}\nu }^{(S)}$ satisfty Eq.(\ref{homHelm}), it is easy to prove that the scattered field is given by
\begin{eqnarray} \label{Fskl}
&& {\bf{F}}_{\omega {\bf{n}}\nu }^{(sc)} \left( {\bf{r}} \right) = -\int {d^3 {\bf{r}}'} {\cal G}_\omega  \left( {{\bf{r}},{\bf{r}}'} \right) \cdot  \nonumber \\ 
&&  \cdot \left[ {  \nabla _{{\bf{r}}'}  \times \frac{1}{{\mu _\omega  \left( {{\bf{r}}'} \right)}}\nabla _{{\bf{r}}'}  \times  - k_\omega^2 \varepsilon _\omega  \left( {{\bf{r}}'} \right)} \right]{\bf{F}}_{\omega {\bf{n}}\nu }^{(in)} \left( {{\bf{r}}'} \right),
 \end{eqnarray}
an expression lucidly clarifying that the dyadic Green's function provides the description of the scattering modes as well, so that its knowledge amounts to a complete description of field-macroscopic matter interaction.  Note that the integrand in the right hand side of Eq.(\ref{Fskl}) vanishes for ${\bf r}'$ outside the sample volume since the incident plane wave satisfies the vacuum Helmholtz equation $\left( {\nabla  \times \nabla  \times  - k_\omega^2 } \right){\bf{F}}_{\omega {\bf{n}}\nu }^{(in)} = 0$ and accordingly the integration domain can be restricted to the volume occupied by the object. From a physical point of view this means that the scattered field ${\bf{F}}_{\omega {\bf{n}}\nu }^{(sc)}$ can be regarded as being  generated by localized sources inside the sample and accordingly, in view of Eqs.(\ref{GreenAsym}), it is evident from Eq.(\ref{Fskl}) that the scattered field asymptotically satisfies the Sommerfeld radiation condition for $r \rightarrow \infty$ expressed by the relations
\begin{eqnarray} \label{fAsym}
 {\bf{F}}_{\omega {\bf{n}}\nu }^{(sc)} \left( {\bf{r}} \right) &=& \frac{{e^{ik_\omega  r} }}{r}{\bf{w}}_{\omega {\bf{n}}\nu }^{(sc)} \left( {o_{\bf r} } \right) + O\left( {\frac{1}{{r^2 }}} \right), \nonumber \\ 
 \nabla  \times {\bf{F}}_{\omega {\bf{n}}\nu }^{(sc)} \left( {\bf{r}} \right) &=& \frac{{e^{ik_\omega  r} }}{r}ik_\omega  {\bf{u}}_{\bf r}  \times {\bf{w}}_{\omega {\bf{n}}\nu }^{(sc)} \left( {o_{\bf r} } \right) + O\left( {\frac{1}{{r^2 }}} \right), \nonumber \\ 
\end{eqnarray}
where ${\bf{w}}_{\omega {\bf{n}}\nu }^{(sc)} \left( {o_{\bf r} } \right)$ is a vector amplitude which is ortogonal to the observation direction
\begin{equation} \label{tranf}
{\bf{u}}_{\bf r}  \cdot {\bf{w}}_{\omega {\bf{n}}\nu }^{(sc)} \left( {o_{\bf r} } \right) = 0.
\end{equation}

The far field behaviors of the scattering mode in Eqs.(\ref{fAsym}) and of the dyadic Green's function in Eqs.(\ref{GreenAsym}) are the key ingredients to prove the rigorous relation (see Appendix D)
\begin{equation} \label{FW}
{\bf{F}}_{\omega {\bf{n}}\nu }^{(S)} \left( {{\bf{r}}} \right) = 4\pi \;{\bf{e}}_{{\bf{n}}\nu }  \cdot {\cal W}_\omega  \left( {o_{ - {\bf{n}}} ,{\bf{r}}} \right),
\end{equation}
which deserves some discussion at this point. First, this relation shows that the overall scattering mode ${\bf{F}}_{\omega {\bf{n}}\nu }^{(S)}$ is directly available if  the asymptotic dyadic amplitude ${\cal W}_\omega$ is known, thus avoiding the task of performing the integration in Eq.(\ref{Fskl}). Second, Eq.(\ref{FW}) admits the simple physical interpretation (see Appendix D)
according to which the scattering mode ${\bf{F}}_{\omega {\bf{n}}\nu }^{(S)}$ can be regarded as the field produced by a point dipole directed along the incident plane wave polarization ${\bf{e}}_{{\bf{n}}\nu }$ and located infinitely far away from the object in the direction $-\bf{n}$, opposite to the incident wave vector, as intuitively expected. In addition, Eq.(\ref{FW}) has a very important consequence namely the integral relation (see Appendix D)
\begin{eqnarray} \label{FFWW}
&&\int {do_{\bf{n}} } \sum\limits_{\nu =1,2}  {{\bf{F}}_{\omega {\bf{n}}\nu }^{(S)} \left( {{\bf{r}} } \right){\bf{F}}_{\omega {\bf{n}}\nu }^{S*} \left( {{\bf{r}}' } \right)}  = \nonumber \\
&& = 16\pi ^2 \int {do } \;  {\cal W}_\omega ^T \left( {o ,{\bf{r}} } \right) \cdot {\cal W}_\omega ^ *  \left( {o ,{\bf{r}}' } \right),
\end{eqnarray}
which is of particular relevance for our investigation since it relates an integral property of the scattering modes in its left hand side to the surface contribution appearing in the fundamental integral relation in Eq.(\ref{FunIntG}).

\section{Bosonic operators}
We aim now at proving that two set of bosonic operators can be introduced to independently describe the medium assisted field and the scattering field. We start by slightly changing the hiterto used notation in order to fit the standard one adopted in the LNF literature and accordingly, for the medium assisted field we consider the electric and magnetic dyadics 
and operators
\begin{eqnarray} \label{notChan}
 {\cal G}_{\omega e} \left( {{\bf{r}},{\bf{r}}'} \right) &=& i\frac{{\omega ^2 }}{{c^2 }}\sqrt {\frac{\hbar }{{\pi \varepsilon _0 }}{\mathop{\rm Im}\nolimits} \left[ {\varepsilon _\omega  \left( {{\bf{r}}'} \right)} \right]} {\cal G}_\omega  \left( {{\bf{r}},{\bf{r}}'} \right), \nonumber \\ 
 {\cal G}_{\omega m} \left( {{\bf{r}},{\bf{r}}'} \right) &=&  - i\frac{\omega }{c}\sqrt {\frac{\hbar }{{\pi \varepsilon _0 }}{\mathop{\rm Im}\nolimits} \left[ {\frac{{ - 1}}{{\mu _\omega  \left( {{\bf{r}}'} \right)}}} \right]} \left[ {{\cal G}_\omega  \left( {{\bf{r}},{\bf{r}}'} \right) \times \mathord{\buildrel{\lower3pt\hbox{$\scriptscriptstyle\leftarrow$}} 
\over \nabla } _{{\bf{r}}'} } \right], \nonumber  \\ 
 {\bf{\hat f}}_{\omega e} \left( {\bf r} \right) &=&  - i\sqrt {\frac{{2\omega }}{\hbar }} {\bf{\hat Z}}_\omega \left( {\bf r} \right), \nonumber  \\ 
 {\bf{\hat f}}_{\omega m} \left( {\bf r} \right) &=&   \sqrt {\frac{{2\omega }}{\hbar }} {\bf{\hat W}}_\omega \left( {\bf r} \right),  
\end{eqnarray}
where the ${\cal G}$ dyadics are related to the ${\cal A}$ ones by $ {\cal G}_{\omega e} \left( {{\bf{r}},{\bf{r}}'} \right) = i\sqrt {\hbar \mu _0 \omega ^2 /\pi } {\cal A}_{\omega e} \left( {{\bf{r}},{\bf{r}}'} \right)$ and ${\cal G}_{\omega m} \left( {{\bf{r}},{\bf{r}}'} \right) =  - i\sqrt {\hbar \mu _0 \omega ^2 /\pi } {\cal A}_{\omega m} \left( {{\bf{r}},{\bf{r}}'} \right)$, whereas for the scattering field we introduce the vectors and operators
\begin{eqnarray}
 {\bf{E}}_{\omega {\bf{n}}\nu } \left( {\bf{r}} \right) &=& \sqrt {\frac{{\hbar \mu _0 \omega ^3 }}{{16\pi ^3 c}}} {\bf{F}}_{\omega {\bf{n}}\nu }^{(S)} \left( {\bf{r}} \right), \nonumber \\ 
 \hat g_{\omega {\bf{n}}\nu }  &=& \sqrt {\frac{{16\pi ^3 c}}{{\hbar \mu _0 \omega ^3 }}} \hat Q_{\omega {\bf{n}}\nu }.
\end{eqnarray}
Using such definitions, the field operator from Eqs.(\ref{EENES}), (\ref{ENo2}) and (\ref{ES})
turns out to be 
\begin{eqnarray} \label{Eomega}
&& {\bf{\hat E}}_\omega  \left( {\bf{r}} \right) = \int {d^3 {\bf{r}}'} \sum\limits_{\lambda} {{\cal G}_{\omega \lambda } \left( {{\bf{r}},{\bf{r}}'} \right) \cdot {\bf{\hat f}}_{\omega \lambda } \left( {{\bf{r}}'} \right)}  +  \nonumber \\ 
&&  + \int {do_{\bf{n}} } \sum\limits_{\nu } {{\bf{E}}_{\omega {\bf{n}}\nu } \left( {\bf{r}} \right)\hat g_{\omega {\bf{n}}\nu } } ,
\end{eqnarray}
whereas the fundamental integral relation from Eq.(\ref{FunIntG}) becomes
\begin{eqnarray} \label{funG}
&& \int {d^3 {\bf{s}}} \sum\limits_{\lambda } {{\cal G}_{\omega \lambda } \left( {{\bf{r}} ,{\bf{s}}} \right) \cdot {\cal G}_{\omega \lambda }^{*T} \left( {{\bf{r}}' ,{\bf{s}}} \right)}  +  \nonumber \\ 
&&  + \frac{{\hbar \mu _0 \omega ^3 }}{{c\pi }}\int {do } {\cal W}_\omega ^T \left( {o ,{\bf{r}}} \right) \cdot {\cal W}_\omega ^* \left( {o ,{\bf{r}}'} \right) = \nonumber \\ 
&&  = \frac{{\hbar \mu _0 \omega ^2 }}{\pi }{\mathop{\rm Im}\nolimits} \left[ {{\cal G}_\omega  \left( {{\bf{r}} ,{\bf{r}}' } \right)} \right]. 
\end{eqnarray}
We emphasize that the field expression in Eq.(\ref{Eomega}) and the fundamental integral relation in Eq.(\ref{funG}) differ from those usually considered in the LNF \cite{Wester}  because of the scattering modes and of the surface term contribution containing the asymptotic amplitude of the dyadic Green's function, respectively. Note that Eq.(\ref{FFWW}) becomes
\begin{eqnarray} \label{EEWW}
&& \int {do_{\bf{n}} } \sum\limits_\nu  {{\bf{E}}_{\omega {\bf{n}}\nu } \left( {\bf{r}} \right){\bf{E}}_{\omega {\bf{n}}\nu }^* \left( {{\bf{r}}'} \right)}  = \nonumber \\ 
&& = \frac{{\hbar \mu _0 \omega ^3 }}{{\pi c}}\int {do } \: {\cal W}_\omega ^T \left( {o ,{\bf{r}}} \right) \cdot {\cal W}_\omega ^ *  \left( {o ,{\bf{r}}'} \right) .
\end{eqnarray}

We now show that ${\bf{\hat f}}_{\omega \lambda } \left( {\bf{r}} \right)$ and $\hat g_{\omega {\bf{n}}\nu }$ are independent bosonic operators or, in other words, that any possile commutation relations between them vanishes except the fundamental ones
\begin{eqnarray} \label{boson}
 \left[ {{\bf{\hat f}}_{\omega \lambda } \left( {\bf{r}} \right),{\bf{\hat f}}_{\omega '\lambda '}^\dag  \left( {{\bf{r}}'} \right)} \right] &=&  \delta \left( {\omega  - \omega '} \right) \delta _{\lambda \lambda '} \delta \left( {{\bf{r}} - {\bf{r}}'} \right) I , \nonumber  \\ 
 \left[ {\hat g_{\omega {\bf{n}}\nu } ,\hat g_{\omega '{\bf{n}}'\nu '}^\dag  } \right] &=&  \delta \left( {\omega  - \omega '} \right)\delta \left( {o_{\bf{n}}  - o_{{\bf{n}}'} } \right)   \delta _{\nu \nu '}, \nonumber \\
\end{eqnarray}
where the delta function with solid angle argument is defined as $\delta \left( {o_{\bf{n}}  - o_{{\bf{n}}'} } \right) = \delta \left( {\theta _{\bf{n}}  - \theta '_{\bf{n}} } \right)\delta \left( {\varphi _{\bf{n}}  - \varphi '_{\bf{n}} } \right)/\sin \theta _{\bf{n}}$. In order to achieve this goal, we adopt the strategy to assume that the bosonic commutation relations of Eq.(\ref{boson}) hold and to show that the canonical commutation relations of Eq.(\ref{CanComRel}) are self-consistently reproduced. As a first basic observation, we note that Eqs.(\ref{Eomega}) and (\ref{boson}) directly yield
\begin{eqnarray} \label{firCOMM}
&& \left[ {{\bf{\hat E}}_\omega  \left( {\bf{r}} \right),{\bf{\hat E}}_{\omega '}^\dag  \left( {{\bf{r}}'} \right)} \right] = \nonumber \\ 
&&  = \delta \left( {\omega  - \omega '} \right)\left\{ {\int {d^3 {\bf{s}}} \sum\limits_\lambda  {{\cal G}_{\omega \lambda } \left( {{\bf{r}},{\bf{s}}} \right) \cdot {\cal G}_{\omega \lambda }^{*T} \left( {{\bf{r}}',{\bf{s}}} \right)}  + } \right. \nonumber \\ 
&&  + \left. {\int {do_{\bf{n}} } \sum\limits_\nu  {{\bf{E}}_{\omega {\bf{n}}\nu } \left( {\bf{r}} \right){\bf{E}}_{\omega {\bf{n}}\nu }^* \left( {{\bf{r}}'} \right)} } \right\}  
\end{eqnarray}
which, by using the fundamental integral relation of Eq.(\ref{funG}) becomes
\begin{eqnarray} \label{EoEoc}
&& \left[ {{\bf{\hat E}}_\omega  \left( {\bf{r}} \right),{\bf{\hat E}}_{\omega '}^\dag  \left( {{\bf{r}}'} \right)} \right] = \nonumber \\ 
&&  = \delta \left( {\omega  - \omega '} \right)\left\{ {\frac{{\hbar \mu _0 \omega ^2 }}{\pi }{\mathop{\rm Im}\nolimits} \left[ {{\cal G}_\omega  \left( {{\bf{r}},{\bf{r}}'} \right)} \right] + } \right. \nonumber  \\ 
&&  + \int {do_{\bf{n}} } \sum\limits_\nu  {{\bf{E}}_{\omega {\bf{n}}\nu } \left( {\bf{r}} \right){\bf{E}}_{\omega {\bf{n}}\nu }^* \left( {{\bf{r}}'} \right)}  + \nonumber  \\ 
&& \left. {  - \frac{{\hbar \mu _0 \omega ^3 }}{{c\pi }}\int {do } {\cal W}_\omega ^T \left( {o ,{\bf{r}}} \right) \cdot {\cal W}_\omega ^* \left( {o_r ,{\bf{r}}'} \right)} \right\} 
\end{eqnarray}
so that, with the help of Eq.(\ref{EEWW}), we get
\begin{equation} \label{CEEcroc}
\left[ {{\bf{\hat E}}_\omega  \left( {\bf{r}} \right),{\bf{\hat E}}_{\omega '}^\dag  \left( {{\bf{r}}'} \right)} \right] = \delta \left( {\omega  - \omega '} \right)\frac{{\hbar \mu _0 \omega ^2 }}{\pi }{\mathop{\rm Im}\nolimits} \left[ {{\cal G}_\omega  \left( {{\bf{r}},{\bf{r}}'} \right)} \right].
\end{equation}
It is crucial to highlight that the commutator in Eq.(\ref{CEEcroc}) has the same expression it has in the LNF with the substantial difference that, in the MLNF we are examining here, it results from the exact balance in Eq.(\ref{EoEoc}) between the scattering mode contribution and the surface term of the fundamental integral equation, as assured by Eq.(\ref{EEWW}). The second observation is that from Eq.(\ref{Eomega}) and Eqs.(\ref{boson}) we readily obtain the commutator
\begin{equation} \label{CEFcroc}
\left[ {{\bf{\hat E}}_\omega  \left( {\bf{r}} \right),{\bf{\hat f}}_{\omega '\lambda }^\dag  \left( {{\bf{r}}'} \right)} \right] = \delta \left( {\omega  - \omega '} \right){\cal G}_{\omega \lambda } \left( {{\bf{r}},{\bf{r}}'} \right)
\end{equation}
which again has the same expression it has in the LNF. Now it is manifest that the six canonical fields in Eq.(\ref{FreqRepFiel}) solely depend  on ${\bf{\hat E}}_\omega$ and ${\bf{\hat f}}_{\omega \lambda }$ and hence any possible commutation relation between the canonical fields is evidently determined by Eqs.(\ref{CEEcroc}) and (\ref{CEFcroc}) only. As observed above, these two commutation relations coincide with those pertaining the LNF and hence the canonical approach of Ref. \cite{Philbi} where canonical commutation relations  hold true. Therefore, we conclude that the canonical commutation relations of Eqs.(\ref{CanComRel}) are automatically satisfied in the here considered MLNF as well.

As a supplement to such formal proof of the equal-time commutation relations  in Eqs.(\ref{CanComRel}) based on the bosonic commutation relations of Eqs.(\ref{boson}), we here show how they can be explicitly deduced from the frequency domain expressions of the fields in Eq.(\ref{FreqRepFiel}) and the commutation relations in Eqs.(\ref{CEEcroc}) and (\ref{CEFcroc}) by means of the general commutation relation of Eq.(\ref{eqtcomm}) in Appendix A. For the vector potential and its canonical momentum we get after some algebra
\begin{eqnarray}
&& \left[ {{\bf{\hat A}}\left( {{\bf{r}},t} \right),{\bf{\hat \Pi }}_A \left( {{\bf{r}}',t} \right)} \right] = i\hbar \int {d^3 {\bf{s}}} \;\delta ^ \bot  \left( {{\bf{r}} - {\bf{s}}} \right) \cdot \nonumber \\ 
&&  \cdot \frac{2}{{\pi c^2 }} {\mathop{\rm Im}\nolimits} \left[ \int\limits_0^\infty  {d\omega } \; \omega {\cal G}_\omega  \left( {{\bf{s}},{\bf{r}}'} \right)\varepsilon _\omega  \left( {{\bf{r}}'} \right) \right],
\end{eqnarray}
which reproduces the first of the canonical commutation relations of Eqs.(\ref{CanComRel}) since the imaginary part of the inner integral is equal to $\frac{{\pi c^2 }}{2}\delta \left( {{\bf{s}} - {\bf{r}}'} \right)I$ as shown in Appendix E (see Eq.(\ref{I1})). For the reservoir field operators and their canonical momenta, after some heavy but straightforward algebra, we obtain
\begin{eqnarray}
&& \left[ {{\bf{\hat X}}^\Omega  \left( {{\bf{r}},t} \right),{\bf{\hat \Pi }}_X^{\Omega '} \left( {{\bf{r}}',t} \right)} \right] = i\hbar \mu _0 \alpha _\Omega  \left( {\bf{r}} \right)\alpha _{\Omega '} \left( {{\bf{r}}'} \right) \cdot \nonumber \\ 
&& \cdot {\mathop{\rm Re}\nolimits} \left\{ {\frac{2}{\pi }\int\limits_0^\infty  {d\omega } \frac{{\omega ^3 {\mathop{\rm Im}\nolimits} \left[ {{\cal G}_\omega  \left( {{\bf{r}},{\bf{r}}'} \right)} \right]}}{{\left[ {\Omega ^2  - \left( {\omega  + i\eta } \right)^2 } \right]\left[ {\Omega '^2  - \left( {\omega  - i\eta } \right)^2 } \right]}} + } \right. \nonumber  \\ 
&&  \left. { + \frac{{\Omega ^2 {\cal G}_\Omega  \left( {{\bf{r}},{\bf{r}}'} \right)}}{{\Omega '^2  - \left( {\Omega  + i\eta } \right)^2 }} + \frac{{\Omega '^2 {\cal G}_{\Omega '} \left( {{\bf{r}},{\bf{r}}'} \right)}}{{\Omega ^2  - \left( {\Omega ' + i\eta } \right)^2 }}} \right\} + \nonumber  \\ 
&&  + i\hbar \delta \left( {\Omega  - \Omega '} \right)\delta \left( {{\bf{r}} - {\bf{r}}'} \right)I,
\end{eqnarray}
and
\begin{eqnarray}
&&  \left[ {{\bf{\hat Y}}^\Omega  \left( {{\bf{r}},t} \right),{\bf{\hat \Pi }}_Y^{\Omega '} \left( {{\bf{r}}',t} \right)} \right] =  - i\hbar \mu _0 \beta _\Omega  \left( {\bf{r}} \right)\beta _{\Omega '} \left( {{\bf{r}}'} \right)\nabla _{\bf{r}}  \times \nonumber  \\ 
&&   \times {\mathop{\rm Re}\nolimits} \left\{ {\frac{2}{\pi }\int\limits_0^\infty  {d\omega } \frac{{\omega {\mathop{\rm Im}\nolimits} \left[ {{\cal G}_\omega  \left( {{\bf{r}},{\bf{r}}'} \right)} \right]}}{{\left[ {\Omega ^2  - \left( {\omega  + i\eta } \right)^2 } \right]\left[ {\Omega '^2  - \left( {\omega  - i\eta } \right)^2 } \right]}} + } \right. \nonumber  \\ 
&&  \left. { + \frac{{{\cal G}_\Omega  \left( {{\bf{r}},{\bf{r}}'} \right)}}{{\Omega '^2  - \left( {\Omega  + i\eta } \right)^2 }} + \frac{{{\cal G}_{\Omega '} \left( {{\bf{r}},{\bf{r}}'} \right)}}{{\Omega ^2  - \left( {\Omega ' + i\eta } \right)^2 }}} \right\} \times \mathord{\buildrel{\lower3pt\hbox{$\scriptscriptstyle\leftarrow$}} 
\over \nabla } _{{\bf{r}}'}  +  \nonumber \\ 
&&  + i\hbar \delta \left( {\Omega  - \Omega '} \right)\delta \left( {{\bf{r}} - {\bf{r}}'} \right)I,
\end{eqnarray}
and they reproduce the second and the third of the canonical commutation relations of Eqs.(\ref{CanComRel}) since the terms with the curly brackets vanish as show in Appendix E (see Eq.(\ref{I2}) and (\ref{I3})). As a single example of commuting fields we evaluate the equal-time commutator between the vector potential and one of the reservoir field, namely
\begin{eqnarray}
&& \left[ {{\bf{\hat A}}\left( {{\bf{r}},t} \right),{\bf{\hat X}}^\Omega  \left( {{\bf{r}}',t} \right)} \right] =  - i\hbar \mu _0 \alpha _\Omega  \left( {{\bf{r}}'} \right)\int {d^3 {\bf{s}}\delta ^ \bot  \left( {{\bf{r}} - {\bf{s}}} \right) \cdot } \nonumber \\ 
&&  \cdot {\mathop{\rm Re}\nolimits} \left\{ {\frac{2}{\pi }\int\limits_0^\infty  {d\omega } \frac{{\omega {\mathop{\rm Im}\nolimits} \left[ {{\cal G}_\omega  \left( {{\bf{s}},{\bf{r}}'} \right)} \right]}}{{\Omega ^2  - \left( {\omega  - i\eta } \right)^2 }} + {\cal G}_\Omega  \left( {{\bf{s}},{\bf{r}}'} \right)} \right\} 
\end{eqnarray}
which is easily seen to vanish due to the vanishing of the curly bracket term, as show in Appendix E (see Eq.(\ref{I4})).

A final remark concerning the commutation relation is in order. In section II, we outlined  that the operators ${\bf{\hat Z}}_\Omega$ and ${\bf{\hat W}}_\Omega$ show up in the general solutions of the homogeneous counterparts of Eqs.(\ref{XYOscil}) and that they can not be arbitrarily suppressed without violating the canonical commutation relations. We are here to prove this statement by noting that, in the absence of the operator ${\bf{\hat Z}}_\Omega$ and ${\bf{\hat W}}_\Omega$, the noise current density ${\bf{\hat J}}^{(M)}$ in Eq.(\ref{JNo})  vanishes and the medium assisted field ${\bf{\hat E}}_\omega ^{(M)}$ disappears from Eq.(\ref{EENES}) thus reducing the field ${\bf{\hat E}}_\omega$ to coincide with its scattering part ${\bf{\hat E}}_\omega ^{(S)}$. Accordingly, the  commutation relation of Eq.(\ref{firCOMM}) becomes
\begin{eqnarray} \label{CommWrong0}
&& \left[ {{\bf{\hat E}}_\omega  \left( {\bf{r}} \right),{\bf{\hat E}}_{\omega '}^\dag  \left( {{\bf{r}}'} \right)} \right] = \nonumber \\ 
&&  = \delta \left( {\omega  - \omega '} \right)\int {do_{\bf{n}} } \sum\limits_\nu  {{\bf{E}}_{\omega {\bf{n}}\nu } \left( {\bf{r}} \right){\bf{E}}_{\omega {\bf{n}}\nu }^* \left( {{\bf{r}}'} \right)}   
\end{eqnarray}
or, by using Eqs.(\ref{EEWW}) and (\ref{funG}),
\begin{eqnarray} \label{CommWrong}
&& \left[ {{\bf{\hat E}}_\omega  \left( {\bf{r}} \right),{\bf{\hat E}}_{\omega '}^\dag  \left( {{\bf{r}}'} \right)} \right] = \nonumber \\ 
&&   = \delta \left( {\omega  - \omega '} \right)\left\{ {\frac{{\hbar \mu _0 \omega ^2 }}{\pi }{\mathop{\rm Im}\nolimits} \left[ {{\cal G}_\omega  \left( {{\bf{r}},{\bf{r}}'} \right)} \right]} \right. +  \nonumber \\ 
&&  \left. { - \int {d^3 {\bf{s}}} \sum\limits_\lambda  {{\cal G}_{\omega \lambda } \left( {{\bf{r}},{\bf{s}}} \right) \cdot {\cal G}_{\omega \lambda }^{*T} \left( {{\bf{r}}',{\bf{s}}} \right)} } \right\}  
\end{eqnarray}
which by no means, in the presence of absorption, reduces to the correct expression of Eq.(\ref{CEEcroc}) which is the only possible one consistent with the canonical commutation relations, as argued above. Note that in the transparent limit, which is strictly rigorous only for vacuum, ${\mathop{\rm Im}\nolimits} \left( \varepsilon _\omega \right) = {\mathop{\rm Im}\nolimits} \left( \mu _\omega \right) = 0$ so that the dyadics ${\cal G}_{\omega e}$ and ${\cal G}_{\omega m}$ of Eqs.(\ref{notChan}) vanish and Eq.(\ref{CommWrong}) reduces to Eq.(\ref{CEEcroc}), this restoring theoretical consistency and indicating that the medium assisted field ${\bf{\hat E}}_\omega ^{(M)}$ plays no physical role when absorbtion is neglected. We conclude that quantum mechanics requires to retain the free oscillating parts of the reservoir field dynamics (solution of the homogeneous counterparts of Eqs.(\ref{XYOscil})) thus postulating the existence of the quantum sources $\hat \rho ^{(M)}$ and $ {\bf{\hat J}}^{(M)}$ which are essential in the presence of matter dispersion and absorption.

\section{Hamiltonian and polaritons}
We now show that the Hamiltonian is diagonalized by the introduction of the medium assisted  ${\bf{\hat f}}_{\omega \lambda } \left( {\bf{r}} \right)$ and scattering $\hat g_{\omega {\bf{n}}\nu }$ polariton bosonic operators. To this end, we start by eliminating from the Hamiltonian the vector potential ${{\bf{\hat A}}}$, its canonical momentum ${\bf{\hat \Pi }}_A$ and the polarization fields ${\int\limits_0^{ + \infty } {d\Omega } \;\alpha _\Omega  {\bf{\hat X}}^\Omega  }$, ${\int\limits_0^{ + \infty } {d\Omega } \;\beta _\Omega  {\bf{\hat Y}}^\Omega  }$  by inserting the second of Eqs.(\ref{EB}), Eq.(\ref{PA}) and Eqs.(\ref{intFre}) into Eq.(\ref{Ham1}) thus getting
\begin{eqnarray} \label{Ham2}
&& \hat H = \int {d^3 {\bf{r}}} \left[ {\frac{1}{2}\varepsilon _0 {\bf{\hat E}} \cdot {\bf{\hat E}} - \frac{1}{{2\mu _0 }}{\bf{\hat B}} \cdot {\bf{\hat B}} + } \right. \nonumber \\ 
&& \left. { + \left( {{\bf{\hat H}} - {\bf{\hat M}}^{(M)} } \right) \cdot {\bf{\hat B}} + \hat h_R } \right],
\end{eqnarray}
where
\begin{equation}
\hat h_R  = \frac{1}{2}\int\limits_0^{ + \infty } {d\Omega } \left[ {\left( {{\bf{\hat \Pi }}_X^{\Omega 2}  + {\bf{\hat \Pi }}_Y^{\Omega 2} } \right) + \Omega ^2 \left( {{\bf{\hat X}}^{\Omega 2}  + {\bf{\hat Y}}^{\Omega 2} } \right)} \right]
\end{equation}
is the reservoir Hamiltonian density. Here we have also made use of the commutativity between electric field and reservoir field, i.e. $ [ {{\bf{\hat E}}\left( {{\bf{r}},t} \right),{\bf{\hat X}}^\Omega  \left( {{\bf{r}}',t} \right)}  ] = 0$. The reservoir fields ${\bf \hat X}^\Omega$ and ${\bf \hat Y}^\Omega$ and their canonical momenta ${\bf \hat \Pi }_X^{\Omega }$ and ${\bf \hat \Pi }_Y^{\Omega }$ can be eliminated from the reservoir Hamiltonian density $\hat h_R$ by using the causal time-domain expressions of Eqs.(\ref{XOYO}). By doing this, as detailed in Appendix F, the reservoir Hamiltonian density $\hat h_R$ turns out to display a contribution which exactly cancels the term in square bracket in Eq.(\ref{Ham2}) so that the Hamiltonian can be written as
\begin{equation}
\hat H = \hat H^{(M)}  + \hat H^{(S)} 
\end{equation}
where
\begin{equation} \label{HN}
\hat H^{(M)}  = \int\limits_0^{ + \infty } {d\omega } \;\hbar \omega \sum\limits_\lambda  {\int {d^3 {\bf{r}}} \;{\bf{\hat f}}_{\omega \lambda }^\dag   \cdot {\bf{\hat f}}_{\omega \lambda } } 
\end{equation}
and
\begin{eqnarray} \label{HS0}
&& \hat H^{(S)}  = \frac{1}{2}\int\limits_{ - \infty }^t {d\tau } \int {d^3 {\bf{r}}} \cdot \nonumber \\
&& \cdot \left[ {{\bf{\hat E}} \cdot \frac{{\partial \left( {{\bf{\hat D}} + {\bf{\hat P}}^{(M)} } \right)}}{{\partial \tau }} + \frac{{\partial \left( {{\bf{\hat D}} + {\bf{\hat P}}^{(M)} } \right)}}{{\partial \tau }} \cdot {\bf{\hat E}} + } \right. \nonumber \\ 
&& \left. { + \frac{{\partial {\bf{\hat B}}}}{{\partial \tau }} \cdot \left( {{\bf{\hat H}} - {\bf{\hat M}}^{(M)} } \right) + \left( {{\bf{\hat H}} - {\bf{\hat M}}^{(M)} } \right) \cdot \frac{{\partial {\bf{\hat B}}}}{{\partial \tau }}} \right].
\end{eqnarray}
The operator $\hat H^{(M)}$ coincides with the Hamiltonian ruling the LNF (and the macroscopic quantum electrodynamics) and we evidently refer to it as the medium assisted contribution to the Hamiltonian. On the other hand the operator $\hat H^{(S)}$ does not show up in the LNF and we have labelled it with a superscipt $(S)$ since, as we are showing in a while, it is the contribution to the Hamiltonian which is provided by te scattering modes. 

To discuss this fundamental point, we note that the third and fourth quantum Maxwell equation, the second of Eqs.(\ref{homMax}) and the second of Eqs.(\ref{QMeq}) respectively, with the help of the noise current density ${\bf{\hat J}}^{(M)}$ in the second of Eqs.(\ref{NoiSou}), can be written as
\begin{eqnarray}
 \frac{{\partial {\bf{\hat B}}}}{{\partial t}} &=&  - \nabla  \times {\bf{\hat E}}, \nonumber \\ 
 \frac{{\partial \left( {{\bf{\hat D}} + {\bf{\hat P}}^{(M)} } \right)}}{{\partial t}} &=& \nabla  \times \left( {{\bf{\hat H}} - {\bf{\hat M}}^{(M)} } \right), 
\end{eqnarray}
so that they can be used to eliminate the time derivatives in Eq.(\ref{HS0}) thus getting
\begin{eqnarray}
&& \hat H^{(S)}  = \frac{1}{2}\int\limits_{ - \infty }^t {d\tau } \int {d^3 {\bf{r}} \cdot } \nonumber \\ 
&&  \cdot \left\{ {{\bf{\hat E}} \cdot \left[ {\nabla  \times \left( {{\bf{\hat H}} - {\bf{\hat M}}^{(M)} } \right)} \right] - \left( {\nabla  \times {\bf{\hat E}}} \right) \cdot \left( {{\bf{\hat H}} - {\bf{\hat M}}^{(M)} } \right) + } \right. \nonumber \\ 
&& \left. { - \left( {{\bf{\hat H}} - {\bf{\hat M}}^{(M)} } \right) \cdot \left( {\nabla  \times {\bf{\hat E}}} \right) + \left[ {\nabla  \times \left( {{\bf{\hat H}} - {\bf{\hat M}}^{(M)} } \right)} \right] \cdot {\bf{\hat E}}} \right\}. \nonumber \\ 
\end{eqnarray}
Using twice the fundamental operator relation of Eq. (\ref{ProtQop}) proven in Appendix A we readily get
\begin{eqnarray} \label{HS1}
&& \hat H^{(S)}  =  - \frac{1}{2}\int\limits_{ - \infty }^t {d\tau } \int\limits_{S_\infty  } {dS} \;{\bf{u}}_r  \cdot \nonumber \\
&& \cdot \left[ {{\bf{\hat E}} \times \left( {{\bf{\hat H}} - {\bf{\hat M}}^{(M)} } \right) - \left( {{\bf{\hat H}} - {\bf{\hat M}}^{(M)} } \right) \times {\bf{\hat E}}} \right]
\end{eqnarray}
so that, bearing in mind that the sample is  embedded in vacuum in such a way that ${\bf{\hat M}}_N  = 0$ and ${\bf{\hat H}} = \frac{1}{{\mu _0 }}{\bf{\hat B}}$ on ${S_\infty  }$ (i.e. for $r \rightarrow \infty$), Eq.(\ref{HS1}) can be written as
\begin{equation} \label{HS2}
\hat H^{(S)}  =  - \int\limits_{ - \infty }^t {d\tau } \int\limits_{S_\infty  } {dS} \;{\bf{u}}_r  \cdot \left( {\frac{{{\bf{\hat E}} \times {\bf{\hat B}} - {\bf{\hat B}} \times {\bf{\hat E}}}}{{2\mu _0 }}} \right).
\end{equation}
A discussion is here in order since this expression of $\hat H^{(S)}$ admits a simple and important physical interpretation. The vector operator appearing inside the bracket is noting more than the quantum Poynting vector which correctly shows up opportunely symmetrized due the hermitian requirement. Accordingly, the opposite of the surface integral over $S_\infty$ in Eq.(\ref{HS2}) represents the electromagnetic power coming from infinity (as generated by infinitely far sources) at time $\tau$ and its time integral from $-\infty$ to $t$, which is precisely $\hat H^{(S)}$, represents the total electromagnetic energy which has flown from infinity during this upper bounded time period. In view of such interpretation, it is evident that $\hat H^{(S)}$ is necessarily related solely to the scattering modes since their plane wave part ${\bf{\hat F}}_\omega ^{(in)}$ of Eq.(\ref{Fi}) is the only field contribution able to carry energy in from infinity at any finite time. To see it in more detail, note that all the other field contributions, namely the medium assisted field ${\bf{\hat E}}_\omega ^{(M)}$ of Eq.(\ref{ENo2}) and the scattered field ${\bf{\hat F}}_\omega ^{(sc)}$ of Eq.(\ref{Fskl}), can be regarded as generated by sources localized inside the sample volume and they are assumed to be absent at $\tau = -\infty$. Accordingly the field they produces necessarily vanishes on the surface $S_\infty$  
at any finite time since such fields travels in vacuum at the finite speed of light $c$ and the surface is infinitely far from the sample. In order to prove this statements we consider the far field limit $r \rightarrow \infty$
of the electric field operator of Eq.(\ref{Eomega}), namely
\begin{equation}
{\bf{\hat E}}_\omega  \left( {\bf{r}} \right) = {\bf{\hat E}}_\omega ^{(ra)} \left( {\bf{r}} \right) + {\bf{\hat E}}_\omega ^{(in)} \left( {\bf{r}} \right) + O\left( {\frac{1}{{r^2 }}} \right)
\end{equation}
where 
\begin{eqnarray}
 {\bf{\hat E}}_\omega ^{(ra)} \left( {\bf{r}} \right) &=& \frac{{e^{ik_\omega  r} }}{r}{\bf{\hat T}}_\omega  \left( {o_r } \right), \nonumber \\ 
 {\bf{\hat E}}_\omega ^{(in)} \left( {\bf{r}} \right) &=& \sqrt {\frac{{\hbar \mu _0 \omega ^3 }}{{16\pi ^3 c}}} \int {do_{\bf{n}} } \sum\limits_\nu  {e^{ik_\omega  {\bf{n}} \cdot {\bf{r}}} {\bf{e}}_{{\bf{n}}\nu } \hat g_{\omega {\bf{n}}\nu } }  
\end{eqnarray}
are the asymptotic radiated field and the incident field made up of plane waves coming from infinity. In the radiated field expression, the operator ${{\bf{\hat T}}_\omega  \left( {o_r } \right)}$ is independent on $r$ and it collects the asymptotic contributions of the dyadics ${\cal G}_{\omega e }$ and ${\cal G}_{\omega e }$ by means of the first two of Eqs.(\ref{notChan}) and the first of Eqs.(\ref{GreenAsym}) and of the asymptotic amplitude ${\bf{f}}_{\omega {\bf{n}}\nu }^{(sc)}$ defined in the first of Eqs.(\ref{fAsym}) and pertaining the scattered field.  Now the time-domain expression of the radiated field is
\begin{equation} \label{Erad}
{\bf{\hat E}}^{(ra)} \left( {{\bf{r}},t} \right) = \frac{1}{r}\int\limits_0^\infty  {d\omega } e^{ik_\omega  \left( {r - ct} \right)} {\bf{\hat T}}_\omega  \left( {o_r } \right) + {\rm h.c.}
\end{equation}
which is seen to be composed of solely outgoing spherical waves traveling at speed $c$. The crucial point is here that, at any finite time $t$, we have
\begin{equation}
\mathop {\lim }\limits_{r \to \infty } r{\bf{\hat E}}^{(ra)} \left( {{\bf{r}},t} \right) = 0
\end{equation}
as a consequence of the Riemann-Lebesgue lemma  and this shows that the scattering Hamiltonian of Eq.(\ref{HS2}) gets contribution only from the incident field, i.e.
\begin{equation} \label{HS3}
\hat H^{(S)}  =  - \int\limits_{ - \infty }^t {d\tau } \int\limits_{S_\infty  } {dS} \;{\bf{u}}_r  \cdot \frac{{{\bf{\hat E}}^{\left( {in} \right)}  \times {\bf{\hat B}}^{\left( {in} \right)}  - {\bf{\hat B}}^{\left( {in} \right)}  \times {\bf{\hat E}}^{\left( {in} \right)} }}{{2\mu _0 }}, 
\end{equation}
where
\begin{eqnarray}  \label{EinBin}
&& {\bf{\hat E}}^{(in)} \left( {{\bf{r}},t} \right) = \int\limits_0^\infty  {d\omega } \int {do_{\bf{n}} } \sum\limits_\nu  {\sqrt {\frac{{\hbar \mu _0 \omega ^3 }}{{16\pi ^3 c}}}  \cdot } \nonumber  \\ 
 && \cdot  e^{ik_\omega  \left( {{\bf{n}} \cdot {\bf{r}} - ct} \right)} \hat g_{\omega {\bf{n}}\nu }   {\bf{e}}_{{\bf{n}}\nu } + {\rm h.c.}, \nonumber \\
&& {\bf{\hat B}}^{(in)} \left( {{\bf{r}},t} \right) = \int\limits_0^\infty  {d\omega } \int {do_{\bf{n}} } \sum\limits_\nu  {\sqrt {\frac{{\hbar \mu _0 \omega ^3 }}{{16\pi ^3 c^3 }}}  \cdot } \nonumber  \\ 
&&  \cdot e^{ik_\omega  \left( {{\bf{n}} \cdot {\bf{r}} - ct} \right)} \hat g_{\omega {\bf{n}}\nu }  \left({\bf{n}} \times {\bf{e}}_{{\bf{n}}\nu }\right) + {\rm h.c.}.
\end{eqnarray}
Substituting the incident field of Eqs.(\ref{EinBin}) into the scattering Hamiltonian of Eq.(\ref{HS3}), after some manipulation detailed in Appendix G, we eventually get
\begin{equation}
\hat H^{(S)}  = \int\limits_0^\infty  {d\omega } \hbar \omega \int {do_{\bf{n}} } \sum\limits_\nu  {\hat g_{\omega {\bf{n}}\nu }^\dag  \hat g_{\omega {\bf{n}}\nu } } 
\end{equation}
thus proving that the introduction of the $g_{\omega {\bf{n}}\nu }$ operator diagonalizes the scattering Hamiltonian. We conclude that the Hamiltonian pertaining the here considered MLNF  is
\begin{equation} \label{Hamiltonian}
\hat H = \int\limits_0^{ + \infty } {d\omega } \;\hbar \omega \left[ {\int {d^3 {\bf{r}}} \sum\limits_\lambda  {{\bf{\hat f}}_{\omega \lambda }^\dag   \cdot {\bf{\hat f}}_{\omega \lambda } }  + \int {do_{\bf{n}} } \sum\limits_\nu  {\hat g_{\omega {\bf{n}}\nu }^\dag  \hat g_{\omega {\bf{n}}\nu } } } \right].
\end{equation}

As a final remark, we note that the bosonic commutation relation of Eqs.(\ref{boson}) together with the structure of the Hamiltonian in Eq.(\ref{Hamiltonian}) naturally enable the quasiparticle interpretation where the operators ${\bf{\hat f}}_{\omega \lambda }^\dag$, ${\bf{\hat f}}_{\omega \lambda }$ and $\hat g_{\omega {\bf{n}}\nu }^\dag$, $\hat g_{\omega {\bf{n}}\nu }$ are viewed as creation and annihilation operators of two different kinds of polaritons, the medium assisted and the scattering ones. Medium assisted polaritons play a fundamental role in LNF as well and they are quanta of sources, localized inside the absorbing medium, which supports hybrid matter-field excitations. On the other hand, scattering polaritons are not considered in the LNF, as above comprehensively outlined, and they are quanta of scattering modes in analogy to what happens in vacuum quantum electrodynamics where photos are quanta of free field modes. The physical difference bewteen scattering polaritons and photons of vacuum quantum electrodynamics is that the existence of the former is completely consistent with the sample dispersion and absorption  which are here fully and rigorously accounted for. Scattering polaritons can genuinely be regarded as a generalization of photons to situations where dispersion and absorption can not be neglected. This observation is supported by the fact that in the transparent limit, which is strictly rigorous only for vacuum, ${\mathop{\rm Im}\nolimits} \left( \varepsilon _\omega \right) = {\mathop{\rm Im}\nolimits} \left( \mu _\omega \right) = 0$ so that the dyadics ${\cal G}_{\omega e}$ and ${\cal G}_{\omega m}$ of Eqs.(\ref{notChan}) vanish and the electric field operator accordingly becomes independent on the operators ${\bf{\hat f}}_{\omega \lambda }^\dag$, ${\bf{\hat f}}_{\omega \lambda }$, this meaning that medium assisted polaritons can be neither created nor destroyed, i.e. they have no physical existence. This is consistent with the observation at the end of Sec.IV regarding the physical irrelevance of the medium assisted field ${\bf{\hat E}}_\omega ^{(M)}$ in the transparent limit. Likewise, scattering polaritons do not disappear but conversely reduce to genuine photons as quanta of mode fields pertaining the transparent geometry.

\section{Conclusion}
In summary, we have analytically deduced the modified Langevin noise formalism (MLNF) for the general magnedielectric situation using, as sole staring point, the quantum Maxwell equations in turn canonically derived in Ref.\cite{Philbi} as Heisenberg equations generated by a suitable field-reservoir Hamiltonian. The MLNF provides the quantum electrodynamical description of finite-size samples in vacuum, where matter dispersion and absorption are fully accounted for, and, as we have shown here, it avoids the requirement of taking the lossless limit at the end of the calculations, an unavoidable requirement in the standard LNF. This is possibile since in the MLFN the medium assisted field, describing electromagnetic excitations produced by localized sources, is explicitly separated by the scattering field associated to sources infinitely far from the object. In addition, since vacuum is lossless, it is known that a specific asymptotic surface contribution shows up in the fundamental integral relation for the dyadic Green's function and this term is essential since we have here proved that it exactly compensates the contribution of the scattering field (see Eq.(\ref{EEWW})) in the self-consistent derivation of the electric field commutation relations (see Eqs.(\ref{EoEoc}) and (\ref{CEEcroc})). In this paper we have thoroughly investigated this point in the general magnetodielectric situation and provided the hiterto lacking analytical validation of the MLFN. Our reasoning has been crucially based on a classical electromagnetic relation, Eq.(\ref{FW}), we have here derived and which relates the scattering modes with the far field amplitude of the dyadic Green's function, an electromagnetic relation which, to best of our knowledge, has not been discussed in literature. Such electromagnetic relation is responsible for the above discussed crucial balance between the surface term and scattering field contribution in the electric field commutation relations. Besides, we have shown that creation and annihilation operators can be introduced for both the medium assisted field and the scattering field and we also proved that their introduction diagonalizes the Hamiltonian eventually reducing to the sum of two corresponding harmonic-oscillator like terms. The ensuing quasi-particle picture is that the quantum electromagnetic field is describes by two sets of quanta, the medium assisted polaritons and the scattering polaritons. Polaritons of both kind fully experience the effect of matter dispersion and absorption but their roles and physical natures are quite different as clarified by the transparent limit. Indeed, in the absence of absorption, medium polaritons disappear (thus enlighting their specific and crucial role in the general situation) whereas scattering polaritons reduce to photons of standard quantum electrodynamics.  As a final remark, we stress that expressing the field as a superposition of medium assisted and scattering contributions, in addition of providing the ability to comprise vacuum without taking any lossless limit, could in principle make the MLNF very useful in describing experiments of radiation quantum scattering in the presence of complex structures. In such situations, the ab-initio identification of the scattering modes, related solely to the asymptotic amplitude of the Green's function (Eq.(\ref{FW})), could prove vital in the interpretation of the experimental result due to the possible complexity of the full dyadic Green's function.

\appendix

\section{Vector and dyadic relations}
We label with ${\bf{ab}}$ the dyad formed by the vectors ${\bf{a}}$ and ${\bf{b}}$. The identity dyadic is 
\begin{equation}
I = {\bf{u}}_x {\bf{u}}_x  + {\bf{u}}_y {\bf{u}}_y  + {\bf{u}}_z {\bf{u}}_z,
\end{equation}
where ${\bf{u}}_x$, ${\bf{u}}_y$ and ${\bf{u}}_z$ are cartesian unit vectors. The dyadic transverse and longitudinal delta functions are 
\begin{eqnarray} \label{DeltaTL}
 \delta ^ \bot  \left( {\bf{r}} \right) &=& \frac{1}{{\left( {2\pi } \right)^3 }}\int {d^3 {\bf{k}}} \;e^{i{\bf{k}} \cdot {\bf{r}}} \left( {I - \frac{{{\bf{kk}}}}{{k^2 }}} \right), \nonumber \\ 
 \delta ^\parallel  \left( {\bf{r}} \right) &=& \frac{1}{{\left( {2\pi } \right)^3 }}\int {d^3 {\bf{k}}} \;e^{i{\bf{k}} \cdot {\bf{r}}} \left( {\frac{{{\bf{kk}}}}{{k^2 }}} \right).
 \end{eqnarray}
Accodingly the transverse and longitudinal parts of a vector field ${\bf{F}}\left( {\bf{r}} \right)$ are given by
\begin{eqnarray}
 {\bf{F}}^ \bot  \left( {\bf{r}} \right) &=& \int {d^3 {\bf{r}}'} \delta ^ \bot  \left( {{\bf{r}} - {\bf{r}}'} \right) \cdot {\bf{F}}\left( {{\bf{r}}'} \right), \nonumber \\ 
 {\bf{F}}^\parallel  \left( {\bf{r}} \right) &=& \int {d^3 {\bf{r}}'} \delta ^\parallel  \left( {{\bf{r}} - {\bf{r}}'} \right) \cdot {\bf{F}}\left( {{\bf{r}}'} \right).
\end{eqnarray}
In correspondence of the vector ${\bf{P}} = P_j {\bf{u}}_j$ and the dyadic  ${\cal P} = {\cal P}_{ij} {\bf{u}}_i {\bf{u}}_j$ fields, four kinds of curl can be defined according to
\begin{eqnarray}
 \nabla  \times {\bf{P}} &=& \varepsilon _{ijk} \partial _i P_j {\bf{u}}_k, \nonumber  \\ 
 {\bf{P}} \times \mathord{\buildrel{\lower3pt\hbox{$\scriptscriptstyle\leftarrow$}} 
\over \nabla }  &=& \varepsilon _{jik} \partial _i P_j {\bf{u}}_k, \nonumber  \\ 
 \nabla  \times {\cal P} &=& \varepsilon _{kip} \partial _k {\cal P}_{ij} {\bf{u}}_p {\bf{u}}_j, \nonumber  \\ 
 {\cal P} \times \mathord{\buildrel{\lower3pt\hbox{$\scriptscriptstyle\leftarrow$}} 
\over \nabla }  &=& \varepsilon _{jkp} \partial _k {\cal P}_{ij} {\bf{u}}_i {\bf{u}}_p.
\end{eqnarray}

Since in the present paper we mainly consider a magnetodielectric object placed in vacuum, we are dealing with the very general situation where  dielectric permittivity and magnetic permeability are piecewise continuous functions. Accordingly also the electromagnetic field is generally piecewise continuous and therefore it is worth stressing that, if the divergence is intended in the sense of distributions \cite{Kanwal}, the divergence theorem
\begin{equation} \label{divT}
\int\limits_V {d^3 {\bf{r}}} \; \nabla  \cdot {\bf{F}} = \int\limits_{\partial V} {dS} \;{\bf{n}} \cdot {\bf{F}}
\end{equation}
holds true even when the field ${\bf{F}} ( {\bf r})$ is piecewise continuous over the interior of $V$. After setting ${\bf{F}} = {\bf{P}} \times {\bf{Q}}$ and 
choosing a sphere of radius $R$ as the integration domain $V$ in Eq.(\ref{divT}), in the limit $R\rightarrow \infty$  we get 
\begin{eqnarray} \label{FrotF}
&&\int {d^3 {\bf{r}}} \;{\bf{P}} \cdot \left( {\nabla  \times {\bf{Q}}} \right) = \nonumber \\
&& = \int {d^3 {\bf{r}}} \left( {\nabla  \times {\bf{P}}} \right) \cdot {\bf{Q}} - \int\limits_{S_\infty  } {dS} \;{\bf{u}}_{\bf r} \cdot \left( {{\bf{P}} \times {\bf{Q}}} \right)
\end{eqnarray}
where ${\bf{u}}_{\bf r} = {\bf r}/r$ is the radial unit vectors and the surface integral is performed over the sphere $S_\infty$ of infinite radius by means of the limiting prescription
\begin{equation}
\int\limits_{S_\infty  } {dS} \;f = \mathop {\lim }\limits_{R \to \infty } \int\limits_{S_R} {dS} \;f,
\end{equation}
where $S_R$ is the surface of the sphere of radius $R$. Due to the evident relations $\nabla  \times {\bf{P}} =  - {\bf{P}} \times \mathord{\buildrel{\lower3pt\hbox{$\scriptscriptstyle\leftarrow$}} 
\over \nabla }$ and ${\bf{u}}_{\bf r} \cdot \left( {{\bf{P}} \times {\bf{Q}}} \right) =  - {\bf{P}} \cdot \left( {{\bf{u}}_{\bf r} \times {\bf{Q}}} \right)$, it is straigthforward showing that Eq.(\ref{FrotF}) admits the dyadic generalizations
\begin{eqnarray} \label{DrotF}
&& \int {d^3 {\bf{s}}} \;{\cal P}\left( {{\bf{r}},{\bf{s}}} \right) \cdot \left[ {\nabla _{\bf{s}}  \times {\bf{Q}}\left( {\bf{s}} \right)} \right] =  \nonumber \\ 
&& = \int {d^3 {\bf{s}}} \left[ { - {\cal P}\left( {{\bf{r}},{\bf{s}}} \right) \times \mathord{\buildrel{\lower3pt\hbox{$\scriptscriptstyle\leftarrow$}} 
\over \nabla } _{\bf{s}} } \right] \cdot {\bf{Q}}\left( {\bf{s}} \right) + \nonumber \\ 
&& + \int\limits_{S_\infty  } {dS} \;{\cal P}\left( {{\bf{r}},{\bf{s}}} \right) \cdot \left[ {{\bf{u}}_{\bf s} \times {\bf{Q}}\left( {\bf{s}} \right)} \right] 
\end{eqnarray}
and
\begin{eqnarray} \label{DrotD}
&& \int {d^3 {\bf{s}}} \;{\cal P}\left( {{\bf{r}} ,{\bf{s}}} \right) \cdot \left[ {\nabla _{\bf{s}}  \times {\cal Q}\left( {{\bf{s}},{\bf{r}}' } \right)} \right] = \nonumber \\ 
&& = \int {d^3 {\bf{s}}} \left[ { - {\cal P}\left( {{\bf{r}} ,{\bf{s}}} \right) \times \mathord{\buildrel{\lower3pt\hbox{$\scriptscriptstyle\leftarrow$}} 
\over \nabla _{\bf{s}}  } } \right] \cdot {\cal Q}\left( {{\bf{s}},{\bf{r}}' } \right) + \nonumber \\ 
&& + \int\limits_{S_\infty  } {dS} \;{\cal P}\left( {{\bf{r}} ,{\bf{s}}} \right) \cdot \left[ {{\bf{u}}_{\bf s} \times {\cal Q}\left( {{\bf{s}},{\bf{r}}' } \right)} \right], 
\end{eqnarray}
where ${\cal P}\left( {{\bf{r}},{\bf{r}}'} \right)$ and ${\cal Q}\left( {{\bf{r}},{\bf{r}}'} \right)$ are piecewise continuous dyadic functions.

The commmutator of two field operators ${\bf{\hat P}}= \hat P_i {\bf{u}}_i $ and ${\bf{\hat Q}} = \hat Q_j {\bf{u}}_j$ is the dyadic operator 
\begin{equation}
\left[ {{\bf{\hat P}},{\bf{\hat Q}}} \right] = \left[ {\hat P_i ,\hat Q_j } \right]{\bf{e}}_i {\bf{e}}_j.
\end{equation}
The equal-time commutator of two field operators ${\bf{\hat P}}\left( {{\bf{r}},t} \right)$ and ${\bf{\hat Q}}\left( {{\bf{r}},t} \right)$ admitting the frequency domain representation of Eq.(\ref{HeisenOpe}) is straightforwardly seen to be given by 
\begin{eqnarray} \label{eqtcomm}
&& \left[ {{\bf{\hat P}}\left( {{\bf{r}},t} \right),{\bf{\hat Q}}\left( {{\bf{r}}',t} \right)} \right] = \nonumber \\ 
&&  = \int\limits_0^\infty  {d\omega } \int\limits_0^\infty  {d\omega '} \left\{ {e^{ - i\left( {\omega  + \omega '} \right)t} \left[ {{\bf{\hat P}}_\omega  \left( {\bf{r}} \right),{\bf{\hat Q}}_{\omega '} \left( {{\bf{r}}'} \right)} \right]} \right. + \nonumber  \\ 
&&  + \left. {e^{ - i\left( {\omega  - \omega '} \right)t} \left[ {{\bf{\hat P}}_\omega  \left( {\bf{r}} \right),{\bf{\hat Q}}_{\omega '}^\dag  \left( {{\bf{r}}'} \right)} \right]} \right\} - {\rm h.c.}.
\end{eqnarray}
Since the order of the fields has not been changed in the derivation of Eq.(\ref{FrotF}), it is evident that it holds for field operators as well, i.e.
\begin{eqnarray} \label{ProtQop}
&&\int {d^3 {\bf{r}}} \;{\bf{\hat P}} \cdot \left( {\nabla  \times {\bf{\hat Q}}} \right) = \nonumber \\
&& = \int {d^3 {\bf{r}}} \left( {\nabla  \times {\bf{\hat P}}} \right) \cdot {\bf{\hat Q}} - \int\limits_{S_\infty  } {dS} \;{\bf{u}}_r  \cdot \left( {{\bf{\hat P}} \times {\bf{\hat Q}}} \right). \nonumber \\
\end{eqnarray}

\section{Dyadic Green's function reciprocity and fundamental integral relation}
After relabelling  $\bf r \rightarrow \bf s$ in Eq.(\ref{GreenEQ}), scalar multiplying both of its sides by ${\cal G}_\omega ^T \left( {{\bf{s}},{\bf{r}} } \right)$ and integrating  over $\bf s$, we get
\begin{eqnarray}
&& {\cal G}_\omega ^T \left( {{\bf{r}}' ,{\bf{r}} } \right) = \nonumber \\
&& = \int {d^3 {\bf{s}}} \;{\cal G}_\omega ^T \left( {{\bf{s}},{\bf{r}} } \right) \cdot \left\{ {\nabla_{\bf s}  \times \left[ {\frac{\nabla_{\bf s}  \times {\cal G}_\omega  \left( {{\bf{s}},{\bf{r}}' } \right)}{{\mu _\omega  \left( {\bf{s}} \right)}}} \right]} \right\} + \nonumber \\ 
&& -   k_\omega ^2 \int {d^3 {\bf{s}}} \;\varepsilon _\omega  \left( {\bf{s}} \right){\cal G}_\omega ^T \left( {{\bf{s}},{\bf{r}} } \right) \cdot {\cal G}_\omega  \left( {{\bf{s}},{\bf{r}}' } \right) 
\end{eqnarray}
which, using Eq.(\ref{DrotD}) of Appendix A and the dyadic identity 
\begin{equation} \label{dyaiden0}
{ - {\cal G}_\omega ^T \left( {{\bf{s}},{\bf{r}} } \right) \times \mathord{\buildrel{\lower3pt\hbox{$\scriptscriptstyle\leftarrow$}} 
\over \nabla_{\bf s} } }  = \left[ {\nabla_{\bf s}  \times {\cal G}\left( {{\bf{s}},{\bf{r}} } \right)} \right]^T,
\end{equation}
turns into
\begin{eqnarray} \label{GT21}
&& {\cal G}_\omega ^T \left( {{\bf{r}}' ,{\bf{r}} } \right) = \nonumber \\
&& = \int\limits_{S_\infty  } {dS} \;\frac{{{\cal G}_\omega ^T \left( {{\bf{s}},{\bf{r}} } \right) \cdot \left[ {{\bf{u}}_{\bf s}  \times \nabla_{\bf s}  \times {\cal G}_\omega  \left( {{\bf{s}},{\bf{r}}' } \right)} \right]}}{{\mu _\omega  \left( {\bf{s}} \right)}} +  \nonumber \\ 
&& + \int {d^3 {\bf{s}}} \frac{{\left[ {\nabla_{\bf s}  \times {\cal G}\left( {{\bf{s}},{\bf{r}} } \right)} \right]^T  \cdot \left[ {\nabla_{\bf s}  \times {\cal G}_\omega  \left( {{\bf{s}},{\bf{r}}' } \right)} \right]}}{{\mu _\omega  \left( {\bf{r}} \right)}} + \nonumber \\ 
&& - k_\omega ^2 \int {d^3 {\bf{s}}} \;\varepsilon _\omega  \left( {\bf{s}} \right){\cal G}_\omega ^T \left( {{\bf{s}},{\bf{r}} } \right) \cdot {\cal G}_\omega  \left( {{\bf{s}},{\bf{r}}' } \right).
\end{eqnarray}
It is now very useful exchanging ${\bf r}$ and ${\bf r}'$ in Eq.(\ref{GT21}) and taking its  transpose thus getting
\begin{eqnarray} \label{G12}
&& {\cal G}_\omega  \left( {{\bf{r}} ,{\bf{r}}' } \right) = \nonumber \\
&& \int\limits_{S_\infty  } {dS} \;\frac{{\left[ {{\bf{u}}_{\bf s}  \times \nabla_{\bf s}  \times {\cal G}_\omega  \left( {{\bf{s}},{\bf{r}} } \right)} \right]^T  \cdot {\cal G}_\omega  \left( {{\bf{s}},{\bf{r}}' } \right)}}{{\mu _\omega  \left( {\bf{s}} \right)}} +  \nonumber \\ 
&& + \int {d^3 {\bf{s}}} \frac{{\left[ {\nabla_{\bf s}  \times {\cal G}\left( {{\bf{s}},{\bf{r}} } \right)} \right]^T  \cdot \left[ {\nabla_{\bf s}  \times {\cal G}_\omega  \left( {{\bf{s}},{\bf{r}}' } \right)} \right]}}{{\mu _\omega  \left( {\bf{s}} \right)}} + \nonumber \\ 
&& - k_\omega ^2 \int {d^3 {\bf{s}}} \;\varepsilon _\omega  \left( {\bf{s}} \right){\cal G}_\omega ^T \left( {{\bf{s}},{\bf{r}} } \right) \cdot {\cal G}_\omega  \left( {{\bf{s}},{\bf{r}}' } \right).
\end{eqnarray}
so that the subtraction of Eq.(\ref{GT21}) from Eq.(\ref{G12}) readily yields
\begin{eqnarray} \label{reci0}
&&{\cal G}_\omega  \left( {{\bf{r}} ,{\bf{r}}' } \right) - {\cal G}_\omega ^T \left( {{\bf{r}}' ,{\bf{r}} } \right) = \nonumber \\ 
&& = \int\limits_{S_\infty  } {dS} \;\left\{ {\frac{{\left[ {{\bf{u}}_{\bf s}  \times \nabla_{\bf s}  \times {\cal G}_\omega  \left( {{\bf{s}},{\bf{r}} } \right)} \right]^T  \cdot {\cal G}_\omega  \left( {{\bf{s}},{\bf{r}}' } \right)}}{{\mu _\omega  \left( {\bf{s}} \right)}} + } \right. \nonumber \\ 
&& - \left. {\frac{{{\cal G}_\omega ^T \left( {{\bf{s}},{\bf{r}} } \right) \cdot \left[ {{\bf{u}}_{\bf s}  \times \nabla_{\bf s}  \times {\cal G}_\omega  \left( {{\bf{s}},{\bf{r}}' } \right)} \right]}}{{\mu _\omega  \left( {\bf{s}} \right)}}} \right\}.
\end{eqnarray}
Note that the right hand side of this equation contains surface integral contributions only and hence it can be evaluated by means of the asymptotic behavior of the dyadic Green's function. By using Eqs.(\ref{GreenAsym}) together with the relation
\begin{equation} \label{uxuxW}
{\bf{u}}_{\bf s}  \times \left[ {{\bf{u}}_{\bf s}  \times {\cal W}_\omega  \left( {o_{\bf s} ,{\bf{r}}'} \right)} \right] =  - {\cal W}_\omega  \left( {o_{\bf s} ,{\bf{r}}'} \right)
\end{equation}
which easily follows from Eq.(\ref{tranW}), we  obtain

\begin{eqnarray}
&& \frac{{\left[ {{\bf{u}}_{\bf s}  \times \nabla _{\bf{s}}  \times {\cal G}_\omega  \left( {{\bf{s}},{\bf{r}} } \right)} \right]^T  \cdot {\cal G}_\omega  \left( {{\bf{s}},{\bf{r}}' } \right)}}{{\mu _\omega  \left( {\bf{s}} \right)}} = \nonumber \\ 
&& = - \frac{{e^{i2k_\omega  s} }}{{s^2 }}ik_\omega  {\cal W}_\omega ^T \left( {o_{\bf s} ,{\bf{r}} } \right) \cdot {\cal W}_\omega  \left( {o_{\bf s} ,{\bf{r}}' } \right) + O\left( {\frac{1}{{s^3 }}} \right), \nonumber  \\ 
&& \frac{{{\cal G}_\omega ^T \left( {{\bf{s}},{\bf{r}} } \right) \cdot \left[ {{\bf{u}}_{\bf s}  \times \nabla _{\bf{s}}  \times {\cal G}_\omega  \left( {{\bf{s}},{\bf{r}}' } \right)} \right]}}{{\mu _\omega  \left( {\bf{s}} \right)}} =  \nonumber  \\ 
&& = - \frac{{e^{i2k_\omega  s} }}{{s^2 }}ik_\omega  {\cal W}_\omega ^T \left( {o_{\bf s} ,{\bf{r}} } \right) \cdot {\cal W}_\omega  \left( {o_{\bf s} ,{\bf{r}}'} \right) + O\left( {\frac{1}{{s^3 }}} \right), \nonumber \\
\end{eqnarray}
where we also have used the condition $\mu _\omega = 1$ on $S_\infty$. Inserting these expressions into Eq.(\ref{reci0}) we readily get
\begin{equation}
{\cal G}_\omega  \left( {{\bf{r}} ,{\bf{r}}' } \right) - {\cal G}_\omega ^T \left( {{\bf{r}}' ,{\bf{r}} } \right) = \int\limits_{S_\infty  } {dS} \;O\left( {\frac{1}{{s^3 }}} \right) = 0
\end{equation}
which is the reciprocity relation for the dyadic Green's function.

We now relabel  $\bf r \rightarrow \bf s$ in Eq.(\ref{GreenEQ}) and we scalar multiply both of its sides by ${\cal G}_\omega ^ *  \left( {{\bf{r}} ,{\bf{s}}} \right)$ so that, after integrating over $\bf s$,  we get
\begin{eqnarray} \label{G*1}
&& {\cal G}_\omega ^ *  \left( {{\bf{r}} ,{\bf{r}}' } \right) = \nonumber \\
&& = \int {d^3 {\bf{s}}} \;{\cal G}_\omega ^ *  \left( {{\bf{r}} ,{\bf{s}}} \right) \cdot \left\{ {\nabla _{\bf{s}}  \times \left[ {\frac{{\nabla _{\bf{s}}  \times {\cal G}_\omega  \left( {{\bf{s}},{\bf{r}}' } \right)}}{{\mu _\omega  \left( {\bf{s}} \right)}}} \right]} \right\} +  \nonumber \\ 
&& - k_\omega ^2 \int {d^3 {\bf{s}}} \;\varepsilon _\omega  \left( {\bf{s}} \right){\cal G}_\omega ^ *  \left( {{\bf{r}} ,{\bf{s}}} \right) \cdot {\cal G}_\omega  \left( {{\bf{s}},{\bf{r}}' } \right).
\end{eqnarray}
By using Eq.(\ref{DrotD}) of Appendix A and the dyadic identity 
\begin{equation} \label{rotswi}
\left[ { - {\cal G}_\omega    \left( {{\bf{r}} ,{\bf{s}}} \right) \times \mathord{\buildrel{\lower3pt\hbox{$\scriptscriptstyle\leftarrow$}} 
\over \nabla } _{\bf{s}} } \right] = \left[ {\nabla _{\bf{s}}  \times {\cal G}\left( {{\bf{s}},{\bf{r}} } \right)} \right]^{T}, 
\end{equation}
obtained combinig Eq.(\ref{dyaiden0}) with the reciprocity relation ${\cal G}_\omega ^T \left( {{\bf{s}},{\bf{r}} } \right) = {\cal G}_\omega  \left( {{\bf{r}} ,{\bf{s}}} \right)$, Eq.(\ref{G*1}) yields
\begin{eqnarray} \label{G*2}
&& {\cal G}_\omega ^ *  \left( {{\bf{r}} ,{\bf{r}}' } \right) = \nonumber \\
&& = \int\limits_{S_\infty  } {dS} \;\frac{{{\cal G}_\omega ^ *  \left( {{\bf{r}} ,{\bf{s}}} \right) \cdot \left[ {{\bf{u}}_{\bf s}  \times \nabla _{\bf{s}}  \times {\cal G}_\omega  \left( {{\bf{s}},{\bf{r}}' } \right)} \right]}}{{\mu _\omega  \left( {\bf{s}} \right)}} +  \nonumber \\ 
&& + \int {d^3 {\bf{s}}} \frac{{\left[ {\nabla _{\bf{s}}  \times {\cal G}\left( {{\bf{s}},{\bf{r}} } \right)} \right]^{*T}  \cdot \left[ {\nabla _{\bf{s}}  \times {\cal G}_\omega  \left( {{\bf{s}},{\bf{r}}' } \right)} \right]}}{{\mu _\omega  \left( {\bf{s}} \right)}} + \nonumber \\ 
&& - k_\omega ^2 \int {d^3 {\bf{s}}} \;\varepsilon _\omega  \left( {\bf{s}} \right){\cal G}_\omega ^ *  \left( {{\bf{r}} ,{\bf{s}}} \right) \cdot {\cal G}_\omega  \left( {{\bf{s}},{\bf{r}}' } \right) .
\end{eqnarray}
It is now essential noting that the reciprocity relation can be written as ${\cal G}_\omega  \left( {{\bf{r}} ,{\bf{r}}' } \right) = \left[ {{\cal G}_\omega ^ *  \left( {{\bf{r}}' ,{\bf{r}} } \right)} \right]^{T*}$ which, combined with Eq.(\ref{G*2}), yields
\begin{eqnarray} \label{Go12}
&& {\cal G}_\omega  \left( {{\bf{r}} ,{\bf{r}}' } \right) = \nonumber \\
&& = \int\limits_{S_\infty  } {dS} \;\frac{{\left[ {{\bf{u}}_{\bf s}  \times \nabla _{\bf{s}}  \times {\cal G}_\omega ^ *  \left( {{\bf{s}},{\bf{r}} } \right)} \right]^T  \cdot {\cal G}_\omega  \left( {{\bf{s}},{\bf{r}}' } \right)}}{{\mu _\omega ^* \left( {\bf{s}} \right)}} + \nonumber \\ 
&& + \int {d^3 {\bf{s}}} \frac{{\left[ {\nabla _{\bf{s}}  \times {\cal G}_\omega  \left( {{\bf{s}},{\bf{r}} } \right)} \right]^{*T}  \cdot \left[ {\nabla _{\bf{s}}  \times {\cal G}\left( {{\bf{s}},{\bf{r}}' } \right)} \right]}}{{\mu _\omega ^* \left( {\bf{s}} \right)}} + \nonumber \\ 
&& - k_\omega ^2 \int {d^3 {\bf{s}}} \;\varepsilon _\omega ^ *  \left( {\bf{s}} \right){\cal G}_\omega ^ *  \left( {{\bf{r}} ,{\bf{s}}} \right) \cdot {\cal G}_\omega  \left( {{\bf{s}},{\bf{r}}' } \right).
\end{eqnarray}
Equations (\ref{G*2}) and (\ref{Go12}) direclty imply that 
\begin{eqnarray} \label{ImG12}
&& {\mathop{\rm Im}\nolimits} \left[ {{\cal G}_\omega  \left( {{\bf{r}} ,{\bf{r}}' } \right)} \right] = \nonumber \\ 
&&   = \int\limits_{S_\infty  } {dS} \frac{1}{{2i}} \left\{ {\frac{{{\cal G}_\omega  \left( {{\bf{r}} ,{\bf{s}}} \right) \cdot \left[ {{\bf{u}}_{\bf s}  \times \nabla _{\bf{s}}  \times {\cal G}_\omega ^* \left( {{\bf{s}},{\bf{r}}' } \right)} \right]}}{{\mu _\omega ^ *  \left( {\bf{s}} \right)}}} \right. +  \nonumber\\ 
&&  \left. { - \frac{{\left[ {{\bf{u}}_{\bf s}  \times \nabla _{\bf{s}}  \times {\cal G}_\omega  \left( {{\bf{s}},{\bf{r}} } \right)} \right]^T  \cdot {\cal G}_\omega ^* \left( {{\bf{s}},{\bf{r}}' } \right)}}{{\mu _\omega  \left( {\bf{s}} \right)}}} \right\} + \nonumber \\ 
&&   + \int {d^3 {\bf{s}}} \left\{ {k_\omega ^2 {\mathop{\rm Im}\nolimits} \;\left[ {\varepsilon _\omega  \left( {\bf{s}} \right)} \right]{\cal G}_\omega  \left( {{\bf{r}} ,{\bf{s}}} \right) \cdot {\cal G}_\omega ^ *  \left( {{\bf{s}},{\bf{r}}' } \right) + } \right. \nonumber \\ 
&&  \left. { + {\mathop{\rm Im}\nolimits} \left[ {\frac{{ - 1}}{{\mu _\omega  \left( {\bf{s}} \right)}}} \right]\left[ {\nabla _{\bf{s}}  \times {\cal G}_\omega  \left( {{\bf{s}},{\bf{r}} } \right)} \right]^T  \cdot \left[ {\nabla _{\bf{s}}  \times {\cal G}^ *  \left( {{\bf{s}},{\bf{r}}' } \right)} \right]} \right\} \nonumber \\
\end{eqnarray}
where we have also taken the complex conjugate of both sides to stick to the conventional notation. Again the surface term contribution in the right hand side of Eq.(\ref{ImG12}) can be evaluated by means of the dyadic Green's function asymptotic behavior of Eq.(\ref{GreenAsym}) together with Eq.(\ref{uxuxW}) so that, after some algebra, we get
\begin{eqnarray} \label{BFunIntG}
&& {\mathop{\rm Im}\nolimits} \left[ {{\cal G}_\omega  \left( {{\bf{r}} ,{\bf{r}}' } \right)} \right] =    k_\omega \int {do } \; {\cal W}_\omega ^T \left( {o ,{\bf{r}} } \right) \cdot {\cal W}_\omega ^* \left( {o,{\bf{r}}' } \right) + \nonumber \\ 
&& + \int {d^3 {\bf{s}}} \sum\limits_{\lambda  = e,m} {{\cal A}_{\omega \lambda } \left( {{\bf{r}} ,{\bf{s}}} \right) \cdot {\cal A}_{\omega \lambda }^{*T} \left( {{\bf{r}}' ,{\bf{s}}} \right)}  
\end{eqnarray}
where $o = \left( {\theta ,\varphi  } \right)$, $do = \sin \theta  d\theta  d\varphi $ is the solid angle differential, its integration is performed over the whole solid angle and we have introduced the dyadics
\begin{eqnarray}
 {\cal A}_{\omega e} \left( {{\bf{r}},{\bf{r}}'} \right) &=& k_\omega \sqrt {{\mathop{\rm Im}\nolimits} \left[ {\varepsilon _\omega  \left( {{\bf{r}}'} \right)} \right]} {\cal G}_\omega  \left( {{\bf{r}},{\bf{r}}'} \right), \nonumber \\ 
 {\cal A}_{\omega m} \left( {{\bf{r}},{\bf{r}}'} \right) &=&  \sqrt {{\mathop{\rm Im}\nolimits} \left[ {\frac{{ - 1}}{{\mu _\omega  \left( {{\bf{r}}'} \right)}}} \right]} \left[ {{\cal G}_\omega  \left( {{\bf{r}},{\bf{r}}'} \right) \times \mathord{\buildrel{\lower3pt\hbox{$\scriptscriptstyle\leftarrow$}} 
\over \nabla } _{{\bf{r}}'} } \right]. \nonumber  \\ 
\end{eqnarray}
where use has been made of Eq.(\ref{rotswi}) to define ${\cal A}_{\omega m}$. Equation (\ref{BFunIntG}) is the fundamental integral relation of the dyadic Green's function.

\section{Analytical check of the fundamental integral relation pertaining the vacuum dyadic Green's function}

We here check the fundamental integral relation of Eq.(\ref{FunIntG}) in the case where there is no lossy object, i.e. the whole space is filled by vacuum. The dyadic Green's function of vacuum for ${\bf r} \neq {\bf r'}$  is given by \cite{Chewww}
\begin{eqnarray} \label{GVac}
&& {\cal G}_\omega  \left( {{\bf{r}},{\bf{r}}'} \right) = \frac{{k_\omega  e^{is} }}{{4\pi s}}\left[ {\left( {\frac{{ - 1 + is + s^2 }}{{s^2 }}} \right)I} \right. + \nonumber \\ 
&&  + \left. {\left( {\frac{{3 - 3is - s^2 }}{{s^2 }}} \right)\frac{{{\bf{ss}}}}{{s^2 }}} \right],
\end{eqnarray}
where ${\bf{s}} = k_\omega  \left( {{\bf{r}} - {\bf{r}}'} \right)$ and it displays the asymptotic behavior of Eq.(\ref{GreenAsym}) where
\begin{equation} \label{Wvac}
{\cal W}_\omega  \left( {o,{\bf{r}}'} \right) = \frac{{e^{ - ik_\omega  {\bf{u}}_o  \cdot {\bf{r}}'} }}{{4\pi }}\left( {I - {\bf{u}}_o {\bf{u}}_o } \right).
\end{equation}
where $o = \left( {\theta ,\varphi } \right)$ are the polar angles associated to the unit vector
\begin{equation}
{\bf{u}}_o  = \sin \theta \left (\cos \varphi {\bf{e}}_x  +  \sin \varphi {\bf{e}}_y \right) + \cos \theta {\bf{e}}_z .
\end{equation}
Since for vacuum $\varepsilon_\omega = 1$ and $\mu_\omega =1$, from Eqs.(\ref{AeAm}) we get
\begin{eqnarray} 
 {\cal A}_{\omega e} \left( {{\bf{r}},{\bf{r}}'} \right) &=& 0, \nonumber \\ 
 {\cal A}_{\omega m} \left( {{\bf{r}},{\bf{r}}'} \right) &=& 0,
\end{eqnarray}
so that the first contribution in the left hand side of Eq.(\ref{FunIntG}) vanishes and the fundamental integral relation reduces to 
\begin{equation} \label{Fundvac}
k_\omega  \int {do} \;{\cal W}_\omega ^T \left( {o,{\bf{r}}} \right) \cdot {\cal W}_\omega ^* \left( {o,{\bf{r}}'} \right) = {\mathop{\rm Im}\nolimits} \left[ {{\cal G}_\omega  \left( {{\bf{r}},{\bf{r}}'} \right)} \right].
\end{equation}
To explicily check this relation we evaluate the integral in its left hand side and we start noting that, by using Eq.(\ref{Wvac}), the integral can be written as
\begin{equation}
\int {do} \;{\cal W}_\omega ^T \left( {o,{\bf{r}}} \right) \cdot {\cal W}_\omega ^* \left( {o,{\bf{r}}'} \right) = \frac{1}{{16\pi ^2 }}\int {do} \;e^{ - i{\bf{s}} \cdot {\bf{u}}_o } \left( {I - {\bf{u}}_o {\bf{u}}_o } \right)
\end{equation}
so that, after choosing the polar axis to coincide with the $z$ axis in such a way that 
\begin{equation} \label{ssz}
{{\bf{s}} = s{\bf{e}}_z }
\end{equation}
we get
\begin{eqnarray} \label{IIIIIH1}
&&\int {do} \;{\cal W}_\omega ^T \left( {o,{\bf{r}}} \right) \cdot {\cal W}_\omega ^* \left( {o,{\bf{r}}'} \right) = \nonumber \\
&& =\frac{1}{{16\pi ^2 }}\int\limits_0^\pi  {d\theta } \sin \theta e^{ - is\cos \theta } \int\limits_0^{2\pi } {d\varphi } \;\left( {I - {\bf{u}}_o {\bf{u}}_o } \right).
\end{eqnarray}
By using the integral
\begin{equation}
\int\limits_0^{2\pi } {d\varphi } \;{\bf{u}}_o {\bf{u}}_o  = \pi \left[ {\sin ^2 \theta \left( {{\bf{e}}_x {\bf{e}}_x  + {\bf{e}}_y {\bf{e}}_y } \right) + 2\cos ^2 \theta {\bf{e}}_z {\bf{e}}_z } \right],
\end{equation}
the integration over $\varphi$ in Eq.(\ref{IIIIIH1}) can be straightforwardly perfomed thus getting after some algebra
\begin{eqnarray} \label{IIIIIH2}
&& \int {do} \;{\cal W}_\omega ^T \left( {o,{\bf{r}}} \right) \cdot {\cal W}_\omega ^* \left( {o,{\bf{r}}'} \right) =  \nonumber \\ 
&&  = \left( {I + {\bf{e}}_z {\bf{e}}_z } \right)
\frac{1}{{16\pi }} \int\limits_0^\pi  {d\theta } \;e^{ - is\cos \theta } \sin \theta  +  \nonumber \\ 
&&  + 
\left( {I - 3{\bf{e}}_z {\bf{e}}_z } \right)
\frac{1}{{16\pi }} \int\limits_0^\pi  {d\theta } \;e^{ - is\cos \theta } \sin \theta \cos ^2 \theta.  \nonumber \\ 
\end{eqnarray}
The integrations over $\theta$ are straightforward as well due to the integrals
\begin{eqnarray}
 \int\limits_0^\pi  {d\theta } \;e^{ - is\cos \theta } \sin \theta  &=& \frac{2}{s}\sin \left( s \right), \nonumber \\ 
 \int\limits_0^\pi  {d\theta } \;e^{ - is\cos \theta } \cos ^2 \theta \sin \theta  &=& \frac{4}{{s^2 }}\cos \left( s \right) + \left( {\frac{2}{s} - \frac{4}{{s^3 }}} \right)\sin \left( s \right), \nonumber \\
\end{eqnarray}
so that from Eq.(\ref{IIIIIH2}) we get after some algebra
\begin{eqnarray}
&& k_\omega \int {do} \;{\cal W}_\omega ^T \left( {o,{\bf{r}}} \right) \cdot {\cal W}_\omega ^* \left( {o,{\bf{r}}'} \right) = \nonumber \\ 
&&  = \frac{k_\omega}{{4\pi s}}\left\{ {\frac{{\cos \left( s \right)}}{s}\left( {I - 3{\bf{e}}_z {\bf{e}}_z } \right) + } \right. \nonumber \\ 
&& \left. { + \frac{{\sin \left( s \right)}}{{s^2 }}\left[ {\left( {s^2  - 1} \right)I - \left( {s^2  - 3} \right){\bf{e}}_z {\bf{e}}_z } \right]} \right\}.
\end{eqnarray}
On the other hand, it is immediate to get from Eq.(\ref{GVac}) the relation
\begin{eqnarray}
&& {\mathop{\rm Im}\nolimits} \left[ {{\cal G}_\omega  \left( {{\bf{r}},{\bf{r}}'} \right)} \right] =  \nonumber \\ 
 && = \frac{{k_\omega  }}{{4\pi s}}\left\{ {\frac{{\cos \left( s \right)}}{s}\left( {I - 3\frac{{{\bf{ss}}}}{{s^2 }}} \right) + } \right.  \nonumber \\ 
&& \left. { + \frac{{\sin \left( s \right)}}{{s^2 }}\left[ {\left( {s^2  - 1} \right)I - \left( {s^2  - 3} \right)\frac{{{\bf{ss}}}}{{s^2 }}} \right]} \right\} 
\end{eqnarray}
so that the proof that Eq.(\ref{Fundvac}) is satisfied is complete after noting that Eq.(\ref{ssz}) implies that ${\bf{e}}_z {\bf{e}}_z  = {\bf{ss}}/s^2$.

\section{Relations between scattering modes and dyadic Green's function far field amplitude}
The scattering modes ${\bf{F}}_{\omega {\bf{n}}\nu }^{(S)}$ introduced in Subsec. IIIB are solutions of the homogeneous Helmholtz equation
\begin{equation}
\left( {\nabla  \times \frac{1}{{\mu _\omega  }}\nabla  \times  - k_\omega ^2 \varepsilon _\omega  } \right){\bf{F}}_{\omega {\bf{n}}\nu }^{(S)}  = 0
\end{equation}
and, as a consequence, it can easily be proved that they satisfy the vector Huygens' principle relation \cite{Chewww}
\begin{eqnarray}
&& {\bf{F}}_{\omega {\bf{n}}\nu }^{(S)} \left( {{\bf{r}}'} \right) = \nonumber \\
&& =   - \int\limits_{S_R } {dS} \;{\bf{u}}_{\bf r}  \cdot \left\{ {\frac{{{\bf{F}}_{\omega {\bf{n}}\nu }^{(S)} \left( {\bf{r}} \right) \times \left[ {\nabla  \times {\cal G}_\omega  \left( {{\bf{r}},{\bf{r}}'} \right)} \right]}}{{\mu _\omega  \left( {\bf{r}} \right)}} + } \right. \nonumber \\ 
&& + \left. {  \frac{{\left[ {\nabla  \times {\bf{F}}_{\omega {\bf{n}}\nu }^{(S)} \left( {\bf{r}} \right)} \right] \times {\cal G}_\omega  \left( {{\bf{r}},{\bf{r}}'} \right)}}{{\mu _\omega  \left( {\bf{r}} \right)}}} \right\}
\end{eqnarray}
where the point ${\bf r}'$ lies inside the sphere $S_R $ of radius $R$ over which the integration is carried out. After using the expression of the scattering modes ${\bf{F}}_{\omega {\bf{n}}\nu }^{(S)} = {\bf{F}}_{\omega {\bf{n}}\nu }^{(in)} + {\bf{F}}_{\omega {\bf{n}}\nu }^{(sc)}$ (see Eq.(\ref{FSKl})), we take the limit $R \rightarrow \infty$ of this expression by exploiting the asymptotic behavior of both the Green's function Eq.(\ref{GreenAsym}) and the scattered field Eq.(\ref{fAsym}) so that, after some algebra, we obtain
\begin{equation} \label{FS11}
{\bf{F}}_{\omega {\bf{n}}\nu }^{\left( S \right)} \left( {{\bf{r}}'} \right) = {\bf{U}}_{\omega {\bf{n}}\nu }^{\left( {in} \right)} \left( {{\bf{r}}'} \right) + {\bf{U}}_{\omega {\bf{n}}\nu }^{\left( {sc} \right)} \left( {{\bf{r}}'} \right)
\end{equation}
where
\begin{eqnarray} 
&& {\bf{U}}_{\omega {\bf{n}}\nu }^{(in)} \left( {{\bf{r}}'} \right) =  - \mathop {\lim }\limits_{R \to \infty } R \int {do_{\bf{r}} } e^{ik_\omega  R} {\bf{u}}_{\bf r}  \cdot  \nonumber \\ 
&& \cdot \left\{ {ik_\omega  {\bf{F}}_{\omega {\bf{n}}\nu }^{(in)} \left( {R{\bf{u}}_{\bf r} } \right) \times \left[ {{\bf{u}}_{\bf r}  \times {\cal W}_\omega  \left( {o_{\bf r} ,{\bf{r}}'} \right)} \right] + } \right. \nonumber\\ 
&& +  \left. { \left[ {\nabla  \times {\bf{F}}_{\omega {\bf{n}}\nu }^{(in)} \left( {R{\bf{u}}_{\bf r} } \right)} \right] \times {\cal W}_\omega  \left( {o_{\bf r} ,{\bf{r}}'} \right)} \right\}
\end{eqnarray}
and
\begin{eqnarray}
&& {\bf{U}}_{\omega {\bf{n}}\nu }^{\left( {sc} \right)} \left( {{\bf{r}}'} \right) = 
 - ik_\omega  \mathop {\lim }\limits_{R \to \infty } \int {do_{\bf{r}} } e^{i2k_\omega  R} {\bf{u}}_{\bf r}  \cdot  \nonumber\\ 
&& \cdot  \left\{ {{\bf{w}}_{\omega {\bf{n}}\nu }^{(sc)} \left( {o_{\bf{r}} } \right) \times \left[ {{\bf{u}}_{\bf r}  \times {\cal W}_\omega  \left( {o_{\bf r} ,{\bf{r}}'} \right)} \right]} \right. + \nonumber \\ 
&& + \left. {  \left[ {{\bf{u}}_{\bf r}  \times {\bf{w}}_{\omega {\bf{n}}\nu }^{(sc)} \left( {o_{\bf{r}} } \right)} \right] \times {\cal W}_\omega  \left( {o_{\bf r} ,{\bf{r}}'} \right)} \right\},
\end{eqnarray}
which holds at any point ${\bf r}'$ (we have also used the fact that $\mu_\omega =1$ for $R \rightarrow \infty$). The contribution ${\bf{U}}_{\omega {\bf{n}}\nu }^{\left( {sc} \right)}$  of the scattered field  is easily seen to vanish identically since the relations 
\begin{eqnarray} \label{ABA}
 {\bf{A}} \times \left( {{\bf{B}} \times {\cal A}} \right) &=& {\bf{BA}} \cdot {\cal A} - \left( {{\bf{A}} \cdot {\bf{B}}} \right){\cal A}, \nonumber \\ 
 \left( {{\bf{A}} \times {\bf{B}}} \right) \times {\cal A} &=& {\bf{BA}} \cdot {\cal A} - {\bf{AB}} \cdot {\cal A}
\end{eqnarray}
holding for arbitrary vectors $\bf A$,$\bf B$ and dyadic ${\cal A}$ together with the transversality of both ${\cal W}_\omega$ in Eq.(\ref{tranW}) and ${\bf w}_{\omega {\bf{n}}\nu }^{(sc)}$ in Eq.(\ref{tranf}) readily imply that
\begin{eqnarray}
{{\bf{w}}_{\omega {\bf{n}}\nu }^{(sc)} \left( {o_{\bf{r}} } \right) \times \left[ {{\bf{u}}_{\bf r}  \times {\cal W}_\omega  \left( {o_{\bf r} ,{\bf{r}}'} \right)} \right]} 
  &=& {\bf{u}}_{\bf r}   {\bf{w}}_{\omega {\bf{n}}\nu }^{(sc)} \left( {o_{\bf{r}} } \right) \cdot {\cal W}_\omega  \left( {o_{\bf r} ,{\bf{r}}'} \right), \nonumber\\ 
{\left[ {{\bf{u}}_{\bf r}  \times {\bf{w}}_{\omega {\bf{n}}\nu }^{(sc)} \left( {o_{\bf{r}} } \right)} \right] \times {\cal W}_\omega  \left( {o_{\bf r} ,{\bf{r}}'} \right)} &=&  - {\bf{u}}_{\bf r} {\bf{w}}_{\omega {\bf{n}}\nu }^{(sc)} \left( {o_{\bf{r}} } \right) \cdot {\cal W}_\omega  \left( {o_{\bf r} ,{\bf{r}}'} \right). \nonumber \\ 
\end{eqnarray}
Therefore, by inserting the expression of the plane wave ${\bf{F}}_{\omega {\bf{n}}\nu }^{(in)} \left( {\bf{r}} \right)$ of Eq.(\ref{Fi}) into Eq.(\ref{FS11}), exploiting the relations in Eqs.(\ref{ABA}) together with the transversality of ${\cal W}_\omega$ in Eq.(\ref{tranW}) we get
\begin{eqnarray} \label{FS12}
&& {\bf{F}}_{\omega {\bf{n}}\nu }^{(S)} \left( {{\bf{r}}'} \right) =  - i  \mathop {\lim }\limits_{R \to \infty } (k_\omega R) \int {do_{\bf{r}} } \; e^{i\left( {k_\omega  R} \right)\left( {1 + {\bf{n}} \cdot {\bf{u}}_{\bf r} } \right)}  \cdot  \nonumber \\ 
&& \cdot \left\{ {\left[ {1 - \left( {{\bf{u}}_r  \cdot {\bf{n}}} \right)} \right]{\bf{e}}_{{\bf{n}}\nu }  + {\bf{n}}\left( {{\bf{u}}_r  \cdot {\bf{e}}_{{\bf{n}}\nu } } \right)} \right\} \cdot {\cal W}_\omega  \left( {o_{\bf{r}} ,{\bf{r}}'} \right). \nonumber \\
\end{eqnarray}
To evaluate the limit in the right hand side we resort to the classical Jones' lemma according to which the asymptotic relation 
\begin{eqnarray}
&&  \xi \int {do_{\bf r} } \; e^{i\xi \left( {{\bf{n}} \cdot {\bf{u}}_{\bf r} } \right)} f\left( {o_{\bf r} } \right) = \nonumber \\
&& =2\pi i\left[ {e^{ - i\xi } f\left( {o_{ - {\bf{n}}} } \right) - e^{i\xi } f\left( {o_{\bf{n}} } \right)} \right] + O\left( {\frac{1}{{\xi ^2 }}} \right)
\end{eqnarray}
holds for $\xi \rightarrow \infty$ (see Appendi XII of Ref.\cite{Bornnn}), so that Eq.(\ref{FS12}) readily yields
\begin{equation} \label{FWC}
{\bf{F}}_{\omega {\bf{n}}\nu }^{(S)} \left( {{\bf{r}}} \right) = 4\pi \;{\bf{e}}_{{\bf{n}}\nu }  \cdot {\cal W}_\omega  \left( {o_{ - {\bf{n}}} ,{\bf{r}}} \right),
\end{equation}
where we have relabelled ${\bf r}'$ as $\bf r$.

In addition to its practical utility in providing the scattering modes directly in terms of the far field amplitude of the dyadic Green's function, the rigorous relation in Eq.(\ref{FWC}) admits a very simple physical interpretation. To briefly discuss this point, in correspondence of the scattering mode ${\bf{F}}_{\omega {\bf{n}}\nu }^{(S)}$, consider a point dipole located at 
a distance $\ell$ from the origin along the direction $- {\bf{n}}$ (parallel and opposite to the wave vector of the incident plane wave), whose moment is along the polarization unit vector ${\bf{e}}_{{\bf{n}}\nu } $ and whose strength depends on $\ell$ according to $E_0 \left( {4\pi \varepsilon _0 /k_\omega ^2 } \right)\ell e^{ - ik_\omega  \ell }$, where $E_0$ is a constant dielectric field amplitude. The dipole is described by the current density
\begin{equation}
{\bf{J}}_\omega  \left( {\bf{r}} \right) =  - i\omega \left[ {E_0 \frac{{4\pi \varepsilon _0 }}{{k_\omega ^2 }}\ell e^{ - ik_\omega  \ell } } \right]\delta \left( {{\bf{r}} + \ell {\bf{n}}} \right){\bf{e}}_{{\bf{n}}\nu } 
\end{equation}
which, inserted into the expression
\begin{equation}
{\bf{E}}_\omega  \left( {\bf{r}} \right) = i\omega \mu _0 \int {d^3 {\bf{r}}'} {\cal G}_\omega  \left( {{\bf{r}},{\bf{r}}'} \right) \cdot {\bf{J}}_\omega  \left( {\bf{r}} \right),
\end{equation}
directly yields
\begin{equation}
{\bf{E}}_\omega  \left( {\bf{r}} \right) = 4\pi E_0 \ell e^{ - ik_\omega  \ell } {\cal G}_\omega  \left( {{\bf{r}}, - \ell {\bf{n}}} \right) \cdot {\bf{e}}_{{\bf{n}}\nu } 
\end{equation}
which is the field radiated by the dipole in the presence of the object. We are now to take the limit $\ell \rightarrow \infty$ of this expression and to do so it is convenient to cast is as
\begin{equation}
{\bf{E}}_\omega  \left( {\bf{r}} \right) = 4\pi E_0 \ell e^{ - ik_\omega  \ell } {\bf{e}}_{{\bf{n}}\nu }  \cdot {\cal G}_\omega  \left( { - \ell {\bf{n}},{\bf{r}}} \right)
\end{equation}
where we have used the reciprocity relation of dyadic Green's function (see Eq.(\ref{RecipG})). Exploiting the asymptotic behavior of Eq.(\ref{GreenAsym}), it is straightforward to show that for $\ell \rightarrow \infty$ the precedent relation yieds
\begin{equation}
{\bf{E}}_\omega  \left( {\bf{r}} \right) = 4\pi E_0 {\bf{e}}_{{\bf{n}}\nu }  \cdot {\cal W}_\omega  \left( {o_{ - {\bf{n}}} ,{\bf{r}}} \right),
\end{equation}
which, for $E_0 = 1$, reproduces the scattering mode in Eq.(\ref{FWC}). Therefore we conclude that Eq.(\ref{FWC}) simply states that the scattering mode can be regarded as the field produced by a point dipole directed along the  incident plane wave polarization and located infinitely far away from the object in the opposite direction of the incident wave vector, as intuitively expected.

We now form the dyad ${{\bf{F}}_{\omega {\bf{n}}\nu }^{(S)} \left( {{\bf{r}} } \right){\bf{F}}_{\omega {\bf{n}}\nu }^{S*} \left( {{\bf{r}}' } \right)}$ which, once  integrated over the whole solid angle and summed over the polarizations, yields
\begin{eqnarray} \label{F1F2}
&& \int {do_{\bf{n}} } \sum\limits_\nu  {{\bf{F}}_{\omega {\bf{n}}\nu }^{(S)} \left( {{\bf{r}} } \right){\bf{F}}_{\omega {\bf{n}}\nu }^{S*} \left( {{\bf{r}}' } \right)}  =  \nonumber \\ 
&&  = 16\pi ^2 \int {do_{\bf{n}} } {\cal W}_\omega ^T \left( {o_{ - {\bf{n}}} ,{\bf{r}} } \right) \cdot {\sum\limits_\nu  {{\bf{e}}_{{\bf{n}}\nu } {\bf{e}}_{{\bf{n}}\nu } } }  \cdot {\cal W}_\omega ^ *  \left( {o_{ - {\bf{n}}} ,{\bf{r}}' } \right) \nonumber \\ 
\end{eqnarray}
The polarization unit vectors ${\bf{e}}_{{\bf{n}}1 }$ and ${\bf{e}}_{{\bf{n}}2 }$ together with the wave unit vector $\bf n$ form an orthonormal basis so that
\begin{equation}
\sum\limits_\nu  {{\bf{e}}_{{\bf{n}}\nu } {\bf{e}}_{{\bf{n}}\nu } }  = I - {\bf{nn}}
\end{equation}
and using this relation, together with the transversality of ${\cal W}_\omega$ in Eq.(\ref{tranW}), we cast Eq.(\ref{F1F2}) as
\begin{eqnarray}
&&\int {do_{\bf{n}} } \sum\limits_\nu  {{\bf{F}}_{\omega {\bf{n}}\nu }^{(S)} \left( {{\bf{r}} } \right){\bf{F}}_{\omega {\bf{n}}\nu }^{S*} \left( {{\bf{r}}' } \right)}  = \nonumber \\
&& = 16\pi ^2 \int {do } \;  {\cal W}_\omega ^T \left( {o ,{\bf{r}} } \right) \cdot {\cal W}_\omega ^ *  \left( {o ,{\bf{r}}' } \right).
\end{eqnarray}

\section{Dyadic Green's function integrals}
The Dyadic Green's function is a holomorphic function of $\omega$ in the upper half plane ${\mathop{\rm Im}\nolimits} \; \omega  > 0$ and it is characterized by the reflection principle and the large frequency behavior given by
\begin{eqnarray} \label{GreenPRO}
 {\cal G}_\omega ^ *  \left( {{\bf{r}},{\bf{r}}'} \right) &=& {\cal G}_{ - \omega ^* } \left( {{\bf{r}},{\bf{r}}'} \right), \nonumber  \\ 
 {\cal G}_\omega  \left( {{\bf{r}},{\bf{r}}'} \right) &\approx&  - \frac{{c^2 }}{{\omega ^2 }}\delta \left( {{\bf{r}} - {\bf{r}}'} \right)I, \quad {\rm for} \;\omega  \to \infty,  
\end{eqnarray}
respectively. The first integral we are to evaluate is
\begin{equation}
I_1= {\rm Im} \left[ {\int\limits_0^{ + \infty } {d\omega } \;\omega \,{\cal G}_\omega  \left( {{\bf{r}},{\bf{r}}'} \right)\varepsilon _\omega  \left( {{\bf{r}}'} \right)} \right],
\end{equation}
which, in view of the relation $\left( {{\cal G}_\omega  \varepsilon _\omega  } \right)^*  = {\cal G}_{ - \omega } \varepsilon _{ - \omega } $, can be casted as
\begin{equation}
I_1  = \frac{1}{{2i}}\int\limits_{ - \infty }^{ + \infty } {d\omega } \;\omega \,{\cal G}_\omega  \left( {{\bf{r}},{\bf{r}}'} \right)\varepsilon _\omega  \left( {{\bf{r}}'} \right).
\end{equation}
Since ${\cal G}_\omega$ and $\varepsilon _ \omega$ are holomorphic functions for ${\rm Im } \; \omega >0$, by Cauchy theorem the real axis integration contour can be deformed to a very large semircle in the upper half plane so that
\begin{equation}
I_1 = \frac{1}{2}\mathop {\lim }\limits_{\rho  \to \infty } \rho \int\limits_\pi ^0 {d\theta } e^{i\theta } \left[ {\omega \,{\cal G}_\omega  \left( {{\bf{r}},{\bf{r}}'} \right)\varepsilon _\omega  \left( {{\bf{r}}'} \right)} \right]_{\omega  = \rho e^{i\theta } }, 
\end{equation}
which, by using the the second of Eq.(\ref{GreenPRO}) together with the $\varepsilon _\omega   \approx 1$ for $\omega  \to \infty$, very easily yields
\begin{equation} \label{I1}
{\rm Im} \left[ \int\limits_0^{ + \infty } {d\omega } \;\omega \,{\cal G}_\omega  \left( {{\bf{r}},{\bf{r}}'} \right)\varepsilon _\omega  \left( {{\bf{r}}'} \right)\right] = \frac{{\pi c^2 }}{2}\delta \left( {{\bf{r}} - {\bf{r}}'} \right)I.
\end{equation}
The second integral we need to consider is
\begin{equation}
I_2  = {\mathop{\rm Re}\nolimits} \left\{ {\frac{2}{\pi }\int\limits_0^\infty  {d\omega } \frac{{\omega {\mathop{\rm Im}\nolimits} \left[ {{\cal G}_\omega  \left( {{\bf{r}},{\bf{r}}'} \right)} \right]}}{{\left[ {\Omega ^2  - \left( {\omega  + i\eta } \right)^2 } \right]\left[ {\Omega '^2  - \left( {\omega  - i\eta } \right)^2 } \right]}}} \right\}
\end{equation}
where $\Omega$ and $\Omega'$ are arbitrary positive frequencies. Using the first of Eq.(\ref{GreenPRO}) we get
\begin{eqnarray}
&& I_2  = \frac{1}{{2\pi i}}\int\limits_{ - \infty }^\infty  {d\omega } \; \omega  {{\cal G}_\omega  \left( {{\bf{r}},{\bf{r}}'} \right)}  \cdot \nonumber \\ 
&&  \cdot \left\{ {\frac{1}{{\left[ {\Omega ^2  - \left( {\omega  - i\eta } \right)^2 } \right]\left[ {\Omega '^2  - \left( {\omega  + i\eta } \right)^2 } \right]}} + } \right. \nonumber \\ 
&& + \left. {\frac{1}{{\left[ {\Omega ^2  - \left( {\omega  + i\eta } \right)^2 } \right]\left[ {\Omega '^2  - \left( {\omega  - i\eta } \right)^2 } \right]}}} \right\},
\end{eqnarray}
and we note that the integrand is a holomorphic function in the upper half plane ${\rm Im} \; \omega >0$ except for the four simple poles $\pm \Omega  + i\eta$, $\pm \Omega'  + i\eta$. Since the integrand has asymptotic behavior $1/\omega ^5$ for $\omega  \to \infty$, the integration along the infinite semicircle in the upper half-plane can be  added to close the contour path so that the residue theorem yields
\begin{eqnarray}
 I_2  &=& \frac{{{\cal G}_\Omega  \left( {{\bf{r}},{\bf{r}}'} \right)}}{{2\left[ {\left( {\Omega  + i2\eta } \right)^2  - \Omega '^2 } \right]}} + \frac{{{\cal G}_{ - \Omega } \left( {{\bf{r}},{\bf{r}}'} \right)}}{{2\left[ {\left( {\Omega  - i2\eta } \right)^2  - \Omega '^2 } \right]}} +  \nonumber \\ 
  &+& \frac{{{\cal G}_{\Omega '} \left( {{\bf{r}},{\bf{r}}'} \right)}}{{2\left[ {\left( {\Omega ' + i2\eta } \right)^2  - \Omega ^2 } \right]}} + \frac{{{\cal G}_{ - \Omega '} \left( {{\bf{r}},{\bf{r}}'} \right)}}{{2\left[ {\left( {\Omega ' - i2\eta } \right)^2  - \Omega ^2 } \right]}} 
\end{eqnarray}
or
\begin{eqnarray} \label{I2}
&& {\mathop{\rm Re}\nolimits} \left\{ {\frac{2}{\pi }\int\limits_0^\infty  {d\omega } \frac{{\omega {\mathop{\rm Im}\nolimits} \left[ {{\cal G}_\omega  \left( {{\bf{r}},{\bf{r}}'} \right)} \right]}}{{\left[ {\Omega ^2  - \left( {\omega  + i\eta } \right)^2 } \right]\left[ {\Omega '^2  - \left( {\omega  - i\eta } \right)^2 } \right]}}} \right\} = \nonumber \\ 
&&  =  - {\mathop{\rm Re}\nolimits} \left[ {\frac{{{\cal G}_\Omega  \left( {{\bf{r}},{\bf{r}}'} \right)}}{{\Omega '^2  - \left( {\Omega  + i\eta } \right)^2 }} + \frac{{{\cal G}_{\Omega '} \left( {{\bf{r}},{\bf{r}}'} \right)}}{{\Omega ^2  - \left( {\Omega ' + i\eta } \right)^2 }}} \right], 
\end{eqnarray}
where we have relabelled $2 \eta \rightarrow \eta$ for convenience. The same method enables to prove that
\begin{eqnarray} \label{I3}
&& {\mathop{\rm Re}\nolimits} \left\{ {\frac{2}{\pi }\int\limits_0^\infty  {d\omega } \frac{{\omega ^3 {\mathop{\rm Im}\nolimits} \left[ {{\cal G}_\omega  \left( {{\bf{r}},{\bf{r}}'} \right)} \right]}}{{\left[ {\Omega ^2  - \left( {\omega  + i\eta } \right)^2 } \right]\left[ {\Omega '^2  - \left( {\omega  - i\eta } \right)^2 } \right]}}} \right\} = \nonumber \\ 
&&  =  - {\mathop{\rm Re}\nolimits} \left[ {\frac{{\Omega ^2 {\cal G}_\Omega  \left( {{\bf{r}},{\bf{r}}'} \right)}}{{\Omega '^2  - \left( {\Omega  + i\eta } \right)^2 }} + \frac{{\Omega '^2 {\cal G}_{\Omega '} \left( {{\bf{r}},{\bf{r}}'} \right)}}{{\Omega ^2  - \left( {\Omega ' + i\eta } \right)^2 }}} \right],
\end{eqnarray}
where the only observation is that the integrand has now the asymptotic behavior $1/\omega ^3$ for $\omega \to \infty$ so  that infinite semicircle in the upper half plane with vanishing contribution can again be added to close the integration contour path. Similarly, it is straighforward proving that
\begin{equation} \label{I4}
{\mathop{\rm Re}\nolimits} \left\{ {\frac{2}{\pi }\int\limits_0^\infty  {d\omega } \frac{{\omega {\mathop{\rm Im}\nolimits} \left[ {{\cal G}_\omega  \left( {{\bf{r}},{\bf{r}}'} \right)} \right]}}{{\Omega ^2  - \left( {\omega  - i\eta } \right)^2 }}} \right\} =  - {\mathop{\rm Re}\nolimits} \left[ {{\cal G}_\Omega  \left( {{\bf{r}},{\bf{r}}'} \right)} \right].
\end{equation}

\section{Reservoir Hamiltonian density}
The reservoir Hamiltonian density 
\begin{equation}
\hat h_R  = \frac{1}{2}\int\limits_0^{ + \infty } {d\Omega } \left[ {\left( {{\bf{\hat \Pi }}_X^{\Omega 2}  + {\bf{\hat \Pi }}_Y^{\Omega 2} } \right) + \Omega ^2 \left( {{\bf{\hat X}}^{\Omega 2}  + {\bf{\hat Y}}^{\Omega 2} } \right)} \right]
\end{equation}
can be conveniently rewritten by resorting to the time-dependent expressions of the reservoir fields ${\bf \hat X}^\Omega$ and ${\bf \hat Y}^\Omega$ in Eqs.(\ref{XOYO}) together with those of their conjugated canonical moments ${\bf \hat \Pi }_X^{\Omega }$ and ${\bf \hat \Pi }_Y^{\Omega }$ in the third and fifth Heisenberg equation in Eqs.(\ref{Heisen}), respectively. After some tedious algebra we get
\begin{equation}
\hat h_R  = \hat h_{R1}  + \hat h_{R2}  + \hat h_{R3} 
\end{equation}
where
\begin{eqnarray}
&& \hat h_{R1}  = \int\limits_0^{ + \infty } {d\Omega } \;\Omega ^2 \left( {{\bf{\hat Z}}_\Omega   \cdot {\bf{\hat Z}}_\Omega ^\dag   + {\bf{\hat Z}}_\Omega ^\dag   \cdot {\bf{\hat Z}}_\Omega   + } \right. \nonumber \\ 
&& \left. { + {\bf{\hat W}}_\Omega   \cdot {\bf{\hat W}}_\Omega ^\dag   + {\bf{\hat W}}_\Omega ^\dag   \cdot {\bf{\hat W}}_\Omega  } \right) ,
\end{eqnarray}
\begin{eqnarray}
&& \hat h_{R2}  =  - \frac{i}{2}\int\limits_{ - \infty }^t {d\tau  \cdot } \nonumber \\ 
&& \cdot \left\{ {{\bf{\hat E}}\left( \tau  \right) \cdot \left[ {\int\limits_0^{ + \infty } {d\Omega } \;\Omega \alpha _\Omega  \left( {e^{ - i\Omega \tau } {\bf{\hat Z}}_\Omega   - e^{i\Omega \tau } {\bf{\hat Z}}_\Omega ^\dag  } \right)} \right]} \right. + \nonumber \\ 
&&  + \left[ {\int\limits_0^{ + \infty } {d\Omega } \;\Omega \alpha _\Omega  \left( {e^{ - i\Omega \tau } {\bf{\hat Z}}_\Omega   - e^{i\Omega \tau } {\bf{\hat Z}}_\Omega ^\dag  } \right)} \right] \cdot {\bf{\hat E}}\left( \tau  \right) + \nonumber  \\ 
 && + {\bf{\hat B}}\left( \tau  \right) \cdot \left[ {\int\limits_0^{ + \infty } {d\Omega \;\Omega } \beta _\Omega  \left( {e^{ - i\Omega \tau } {\bf{\hat W}}_\Omega   - e^{i\Omega \tau } {\bf{\hat W}}_\Omega ^\dag  } \right)} \right] +  \nonumber \\ 
&&  + \left. {\left[ {\int\limits_0^{ + \infty } {d\Omega \;} \Omega \beta _\Omega  \left( {e^{ - i\Omega \tau } {\bf{\hat W}}_\Omega   - e^{i\Omega \tau } {\bf{\hat W}}_\Omega ^\dag  } \right)} \right] \cdot {\bf{\hat B}}\left( \tau  \right)} \right\} \nonumber \\ 
\end{eqnarray}
and
\begin{eqnarray} \label{h31}
&&  \hat h_{R3}  = \frac{1}{2}\int\limits_{ - \infty }^t {d\tau } \int\limits_{ - \infty }^t {d\tau '}  \cdot \nonumber \\ 
&&  \cdot \left\{ {{\bf{\hat E}}\left( \tau  \right) \cdot {\bf{\hat E}}\left( {\tau '} \right)\int\limits_0^{ + \infty } {d\Omega }  \; {\alpha _\Omega   ^2} \cos \left[ {\Omega \left( {\tau  - \tau '} \right)} \right] + } \right. \nonumber \\ 
&&  \left. {{\bf{\hat B}}\left( \tau  \right) \cdot {\bf{\hat B}}\left( {\tau '} \right)\int\limits_0^{ + \infty } {d\Omega } \; {\beta _\Omega ^2 }  \,\cos \left[ {\Omega \left( {\tau  - \tau '} \right)} \right]} \right\}.
\end{eqnarray}
By using the definition of the polaritonic operators ${\bf{\hat f}}_{\omega e}$ and ${\bf{\hat f}}_{\omega m}$ in the third and fourth of Eqs.(\ref{notChan}), after relabelling $\Omega$ into $\omega$, the first contribution $\hat h_{R1}$ to the reservoir Hamiltonian density becomes
\begin{equation} 
\hat h_{R1}  = \frac{1}{2}\int\limits_0^{ + \infty } {d\omega } \;\hbar \omega \sum\limits_\lambda  {\left( {\bf{\hat f}}_{\omega \lambda }^\dag   \cdot {\bf{\hat f}}_{\omega \lambda } + {\bf{\hat f}}_{\omega \lambda }  \cdot {\bf{\hat f}}_{\omega \lambda }^\dag\right)} 
\end{equation}
or
\begin{equation} \label{hr1}
\hat h_{R1}  = \int\limits_0^{ + \infty } {d\omega } \;\hbar \omega \sum\limits_\lambda  {{\bf{\hat f}}_{\omega \lambda }^\dag   \cdot {\bf{\hat f}}_{\omega \lambda } }, 
\end{equation}
where use has been made of the boson commutation relation in the first of Eqs.(\ref{boson}) and  the (divergent) zero-point energy has been neglected.

Note now that Eqs.(\ref{PNMN}) containing the definition of the noise polarization ${\bf \hat P}^{(M)}$ and magnetization ${\bf \hat M}^{(M)}$ densities in the frequency domain directly imply that
\begin{eqnarray}
 \frac{{\partial {\bf{\hat P}}^{(M)} }}{{\partial \tau }} &=&  - i\int\limits_0^{ + \infty } {d\Omega } \;\Omega \alpha ^\Omega  \left( {e^{ - i\Omega \tau } {\bf{\hat Z}}_\Omega   - e^{i\Omega \tau } {\bf{\hat Z}}_\Omega ^\dag  } \right), \nonumber \\ 
 \frac{{\partial {\bf{\hat M}}^{(M)} }}{{\partial \tau }} &=&  - i\int\limits_0^{ + \infty } {d\Omega } \,\Omega \beta ^\Omega  \left( {e^{ - i\Omega \tau } {\bf{\hat W}}_\Omega   - e^{i\Omega \tau } {\bf{\hat W}}_\Omega ^\dag  } \right), \nonumber \\
\end{eqnarray}
so that the second contribution $\hat h_{R2}$ to the reservoir Hamiltonian density turns into
\begin{eqnarray} \label{hr2}
&& \hat h_{R2}  = \frac{1}{2}\int\limits_{ - \infty }^t {d\tau } \left[ {{\bf{\hat E}} \cdot \frac{{\partial {\bf{\hat P}}^{(M)} }}{{\partial \tau }} + \frac{{\partial {\bf{\hat P}}^{(M)} }}{{\partial \tau }} \cdot {\bf{\hat E}} + } \right. \nonumber \\ 
&& \left. { + {\bf{\hat B}} \cdot \frac{{\partial {\bf{\hat M}}^{(M)} }}{{\partial \tau }} + \frac{{\partial {\bf{\hat M}}^{(M)} }}{{\partial \tau }} \cdot {\bf{\hat B}}} \right].
\end{eqnarray}
In order to manipulate the third contribution $\hat h_{R3}$ to the reservoir Hamiltonian density we preliminary note that a straightforward calculation yields 
\begin{eqnarray} \label{intab}
&& \int\limits_0^{ + \infty } {d\Omega } \;\alpha _\Omega ^2 ({\bf r}) \cos \left[ {\Omega \left( {\tau  - \tau '} \right)} \right] =  \nonumber \\ 
&&  = \frac{{\varepsilon _0 }}{{2\pi }}\left[ {\frac{{\partial \chi ^{(\varepsilon)}  \left({\bf r}, {\tau  - \tau '} \right)}}{{\partial \tau }} + \frac{{\partial \chi ^{(\varepsilon)}  \left({\bf r}, {\tau ' - \tau } \right)}}{{\partial \tau '}}} \right], \nonumber \\ 
&& \int\limits_0^{ + \infty } {d\Omega \;} \beta _\Omega ^2 ({\bf r}) \cos \left[ {\Omega \left( {\tau  - \tau '} \right)} \right] = \nonumber \\ 
&&  = \frac{1}{{2\pi \mu _0 }}\left[ {\frac{{\partial \chi ^{(\mu)}  \left({\bf r}, {\tau  - \tau '} \right)}}{{\partial \tau }} + \frac{{\partial \chi ^{(\mu)}  \left({\bf r}, {\tau ' - \tau } \right)}}{{\partial \tau '}}} \right],
\end{eqnarray}
where we have used the expressions for the fundamental couping coefficients $\alpha_\omega$ and $\beta_\omega$ in Eqs.(\ref{coupCoeff}), together with the relations $\left( {\chi _\omega ^{\varepsilon ,\mu } } \right)^*  = \chi _{ - \omega }^{\varepsilon ,\mu}$ for the dielectric and magnetic susceptibilities in the frequency domain. Inserting Eq.(\ref{intab}) into Eq.(\ref{h31}) we get
\begin{eqnarray} \label{hr3}
&& \hat h_{R3}  = \frac{1}{2}\int\limits_{ - \infty }^t {d\tau } \left[ {\,{\bf{\hat E}} \cdot \frac{{\partial {\bf{\hat P}}^{(\varepsilon)}  }}{{\partial \tau }} + \frac{{\partial {\bf{\hat P}}^{(\varepsilon)}  }}{{\partial \tau }} \cdot {\bf{\hat E}} + } \right. \nonumber \\ 
&&  \left. { + {\bf{\hat B}} \cdot \frac{{\partial {\bf{\hat M}}^{(\mu)}  }}{{\partial \tau }} + \frac{{\partial {\bf{\hat M}}^{(\mu)}  }}{{\partial \tau }} \cdot {\bf{\hat B}}} \right] 
\end{eqnarray}
where we have used the time-domain expressions of the polarization and magnetization densities of Eq.(\ref{PeMm}). 

By adding the two contributions of Eqs.(\ref{hr2}) and (\ref{hr3}) we get
\begin{eqnarray}
&& \hat h_{R2}  + \hat h_{R3}  =  \nonumber \\ 
&& = \frac{1}{2}\int\limits_{ - \infty }^t {d\tau } \left[ {\,{\bf{\hat E}} \cdot \frac{{\partial \left( {{\bf{\hat P}}^{(\varepsilon)}   + {\bf{\hat P}}^{(M)} } \right)}}{{\partial \tau }} + \frac{{\partial \left( {{\bf{\hat P}}^{(\varepsilon)}   + {\bf{\hat P}}^{(M)} } \right)}}{{\partial \tau }} \cdot {\bf{\hat E}} + } \right. \nonumber \\ 
&& \left. { + {\bf{\hat B}} \cdot \frac{{\partial \left( {{\bf{\hat M}}^{(\mu)}   + {\bf{\hat M}}^{(M)} } \right)}}{{\partial \tau }} + \frac{{\partial \left( {{\bf{\hat M}}^{(\mu)}   + {\bf{\hat M}}^{(M)} } \right)}}{{\partial \tau }} \cdot {\bf{\hat B}}} \right],
\end{eqnarray}
which, using the operator identity
\begin{eqnarray}
&& \varepsilon _0 {\bf{\hat E}} \cdot {\bf{\hat E}} - \frac{1}{{\mu _0 }}{\bf{\hat B}} \cdot {\bf{\hat B}} = \nonumber \\
&& = \int\limits_{ - \infty }^t {d\tau } \left[ {\varepsilon _0 \left( {{\bf{\hat E}} \cdot \frac{{\partial {\bf{\hat E}}}}{{\partial \tau }} + \frac{{\partial {\bf{\hat E}}}}{{\partial \tau }} \cdot {\bf{\hat E}}} \right) + } \right. \nonumber \\ 
&& \left. { - \frac{1}{{\mu _0 }}\left( {{\bf{\hat B}} \cdot \frac{{\partial {\bf{\hat B}}}}{{\partial \tau }} + \frac{{\partial {\bf{\hat B}}}}{{\partial \tau }} \cdot {\bf{\hat B}}} \right)} \right].
\end{eqnarray}
and the definition of the electric displacement ${\bf{\hat D}}$ and magnetic field ${\bf{\hat H}}$ in Eqs.(\ref{DH}), can be casted as
\begin{eqnarray}
&& \hat h_{R2}  + \hat h_{R3}  =  - \frac{1}{2}\varepsilon _0 {\bf{\hat E}} \cdot {\bf{\hat E}} + \frac{1}{{2\mu _0 }}{\bf{\hat B}} \cdot {\bf{\hat B}} + \nonumber  \\ 
&&  + \frac{1}{2}\int\limits_{ - \infty }^t {d\tau } \left[ {{\bf{\hat E}} \cdot \frac{{\partial \left( {{\bf{\hat D}} + {\bf{\hat P}}^{(M)} } \right)}}{{\partial \tau }} + \frac{{\partial \left( {{\bf{\hat D}} + {\bf{\hat P}}^{(M)} } \right)}}{{\partial \tau }} \cdot {\bf{\hat E}} + } \right. \nonumber \\ 
&& \left. { - {\bf{\hat B}} \cdot \frac{{\partial \left( {{\bf{\hat H}} - {\bf{\hat M}}^{(M)} } \right)}}{{\partial \tau }} - \frac{{\partial \left( {{\bf{\hat H}} - {\bf{\hat M}}^{(M)} } \right)}}{{\partial \tau }} \cdot {\bf{\hat B}}} \right].
\end{eqnarray}
After integrating by parts the magnetic integral terms we get
\begin{eqnarray}
&& \hat h_{R2}  + \hat h_{R3}  =  - \frac{1}{2}\varepsilon _0 {\bf{\hat E}} \cdot {\bf{\hat E}} + \frac{1}{{2\mu _0 }}{\bf{\hat B}} \cdot {\bf{\hat B}} +  \nonumber \\ 
&&  - \frac{1}{2} {\bf{\hat B}} \cdot \left( {{\bf{\hat H}} -  {\bf{\hat M}}^{(M)} } \right) - \frac{1}{2} \left( {{\bf{\hat H}} - {\bf{\hat M}}^{(M)} } \right) \cdot {\bf{\hat B}} +   \nonumber \\ 
&&  + \frac{1}{2}\int\limits_{ - \infty }^t {d\tau } \left[ {{\bf{\hat E}} \cdot \frac{{\partial \left( {{\bf{\hat D}} + {\bf{\hat P}}^{(M)} } \right)}}{{\partial \tau }} + \frac{{\partial \left( {{\bf{\hat D}} + {\bf{\hat P}}^{(M)} } \right)}}{{\partial \tau }} \cdot {\bf{\hat E}} + } \right.  \nonumber \\ 
&& \left. { + \frac{{\partial {\bf{\hat B}}}}{{\partial \tau }} \cdot \left( {{\bf{\hat H}} - {\bf{\hat M}}^{(M)} } \right) + \left( {{\bf{\hat H}} - {\bf{\hat M}}^{(M)} } \right) \cdot \frac{{\partial {\bf{\hat B}}}}{{\partial \tau }}} \right],
\end{eqnarray}
so that, adding the contribution $\hat h_{R1}$ of Eq.(\ref{hr1}), the overal reservoir Hamiltonian density turns out to be
\begin{eqnarray}
&& \hat h_R  = \int\limits_0^{ + \infty } {d\omega } \;\hbar \omega \sum\limits_\lambda  {{\bf{\hat f}}_{\omega \lambda }^\dag   \cdot {\bf{\hat f}}_{\omega \lambda } }  + \nonumber \\ 
&&  - \frac{1}{2}\varepsilon _0 {\bf{\hat E}} \cdot {\bf{\hat E}} + \frac{1}{{2\mu _0 }}{\bf{\hat B}} \cdot {\bf{\hat B}} - \left( {{\bf{\hat H}} - {\bf{\hat M}}^{(M)} } \right) \cdot {\bf{\hat B}} + \nonumber \\ 
&&  + \frac{1}{2}\int\limits_{ - \infty }^t {d\tau } \left[ {{\bf{\hat E}} \cdot \frac{{\partial \left( {{\bf{\hat D}} + {\bf{\hat P}}^{(M)} } \right)}}{{\partial \tau }} + \frac{{\partial \left( {{\bf{\hat D}} + {\bf{\hat P}}^{(M)} } \right)}}{{\partial \tau }} \cdot {\bf{\hat E}} + } \right. \nonumber\\ 
&& \left. { + \frac{{\partial {\bf{\hat B}}}}{{\partial \tau }} \cdot \left( {{\bf{\hat H}} - {\bf{\hat M}}^{(M)} } \right) + \left( {{\bf{\hat H}} - {\bf{\hat M}}^{(M)} } \right) \cdot \frac{{\partial {\bf{\hat B}}}}{{\partial \tau }}} \right],  
\end{eqnarray}
where we have also exploited the commutativity between the fields $({\bf{\hat H}} - {\bf{\hat M}}^{(M)})$ and ${\bf{\hat B}}$ in turn resulting from the relation
\begin{equation}
{\bf{\hat H}} - {\bf{\hat M}}^{(M)}  = \frac{1}{{\mu _0 }}{\bf{\hat B}} - \int\limits_0^{ + \infty } {d\Omega } \;\beta _\Omega  {\bf{\hat Y}}^\Omega  
\end{equation}
(see the seconds of Eqs.(\ref{intFre}) and
(\ref{DH})) and the commutativity between magnetic induction field and reservoir field, i.e. $\left[ {{\bf{\hat B}}\left( {{\bf{r}},t} \right),{\bf{\hat Y}}^\Omega  \left( {{\bf{r}}',t} \right)} \right] = 0$.

\section{Scattering Hamiltonian}
In order to manipulate the scattering Hamiltonian $\hat H^{(S)}$ in Eq.(\ref{HS3}) we start by noting that Eq.(\ref{ProtQop}) of Appendix A implies that
\begin{eqnarray} \label{F1}
&& - \int\limits_{S_\infty  } {dS} \;{\bf{u}}_{\bf r} \cdot \left( {{\bf{\hat E}}^{\left( {in} \right)}  \times {\bf{\hat B}}^{\left( {in} \right)}  - {\bf{\hat B}}^{\left( {in} \right)}  \times {\bf{\hat E}}^{\left( {in} \right)} } \right) = \nonumber \\ 
&&  = \int {d^3 {\bf{r}}} \left[ {{\bf{\hat E}}^{\left( {in} \right)}  \cdot \left( {\nabla  \times {\bf{\hat B}}^{\left( {in} \right)} } \right) + \left( {\nabla  \times {\bf{\hat B}}^{\left( {in} \right)} } \right) \cdot {\bf{\hat E}}^{\left( {in} \right)}  + } \right.\nonumber \\ 
&& \left. { - {\bf{\hat B}}^{\left( {in} \right)}  \cdot \left( {\nabla  \times {\bf{\hat E}}^{\left( {in} \right)} } \right) - \left( {\nabla  \times {\bf{\hat E}}^{\left( {in} \right)} } \right) \cdot {\bf{\hat B}}^{\left( {in} \right)} } \right].
\end{eqnarray}
The curls of the electric and magnetic induction fields can be eliminated from this expression after noting that the incident electromagnetic field is a freely propagating field in vacuum and hence it satisfies the vacuum Maxwell equations
\begin{eqnarray}
 \nabla  \times {\bf{\hat E}}^{\left( {in} \right)}  &=&  - \frac{{\partial {\bf{\hat B}}^{\left( {in} \right)} }}{{\partial t}}, \nonumber \\ 
 \nabla  \times {\bf{\hat B}}^{\left( {in} \right)}  &=& \varepsilon _0 \mu _0 \frac{{\partial {\bf{\hat E}}^{\left( {in} \right)} }}{{\partial t}},
\end{eqnarray}
as it can easily verified by means of the expressions in Eq.(\ref{EinBin}). Accordingly from Eq.(\ref{F1}) we get
\begin{eqnarray} \label{F3}
&& - \int\limits_{S_\infty  } {dS} \;{\bf{u}}_{\bf r}  \cdot \left( {\frac{{{\bf{\hat E}}^{\left( {in} \right)}  \times {\bf{\hat B}}^{\left( {in} \right)}  - {\bf{\hat B}}^{\left( {in} \right)}  \times {\bf{\hat E}}^{\left( {in} \right)} }}{{2\mu _0 }}} \right) = \nonumber \\ 
&& = \frac{\partial }{{\partial t}}\int {d^3 {\bf{r}}} \left( {\frac{1}{2}\varepsilon _0 {\bf{\hat E}}^{\left( {in} \right)}  \cdot {\bf{\hat E}}^{\left( {in} \right)}  + \frac{1}{{2\mu _0 }}{\bf{\hat B}}^{\left( {in} \right)}  \cdot {\bf{\hat B}}^{\left( {in} \right)} } \right), \nonumber \\ 
\end{eqnarray}
a very reasonable relation equating the power carried by the incident field and flowing from infinity with the rate of changing of the total electromagnetic energy stored throughout the space by the incident field. By inserting Eq.(\ref{F3}) into Eq.(\ref{HS3}) we find 
\begin{equation} \label{F4}
\hat H^{(S)}  = \int {d^3 {\bf{r}}} \left( {\frac{1}{2}\varepsilon _0 {\bf{\hat E}}^{\left( {in} \right)}  \cdot {\bf{\hat E}}^{\left( {in} \right)}  + \frac{1}{{2\mu _0 }}{\bf{\hat B}}^{\left( {in} \right)}  \cdot {\bf{\hat B}}^{\left( {in} \right)} } \right)
\end{equation}
which states that the scattering Hamiltonian coincides with the electromagnetic energy of the overall incident field, as expected. To proceed, it is convenient setting ${\bf{k}} = k_\omega  {\bf{n}}$ for the wave vector of the plane waves so that Eqs.(\ref{EinBin}) can be written as
\begin{eqnarray}
&& {\bf{\hat E}}^{\left( {in} \right)} \left( {{\bf{r}},t} \right) = c\sqrt {\frac{\hbar }{{16\pi ^3 \varepsilon _0 }}} \int {d^3 {\bf{k}}} \frac{1}{{\sqrt k }}\sum\limits_\nu {\bf{e}}_{{\bf{n}}\nu } \cdot \nonumber  \\ 
&& \cdot \left[ {e^{i\left( {{\bf{k}} \cdot {\bf{r}} - \omega t} \right)} \hat g_{\omega {\bf{n}}\nu }  + e^{ - i\left( {{\bf{k}} \cdot {\bf{r}} - \omega t} \right)} \hat g_{\omega {\bf{n}}\nu }^\dag  } \right], \nonumber \\ 
&& {\bf{\hat B}}^{\left( {in} \right)} \left( {{\bf{r}},t} \right) = c\sqrt {\frac{{\hbar \mu _0 }}{{16\pi ^3 }}} \int {d^3 {\bf{k}}} \frac{1}{{\sqrt k }}\sum\limits_\nu \left( {{\bf{n}} \times {\bf{e}}_{{\bf{n}}\nu } } \right) \cdot \nonumber  \\ 
&& \cdot \left[ {e^{i\left( {{\bf{k}} \cdot {\bf{r}} - \omega t} \right)} \hat g_{\omega {\bf{n}}\nu }  + e^{ - i\left( {{\bf{k}} \cdot {\bf{r}} - \omega t} \right)} \hat g_{\omega {\bf{n}}\nu }^\dag  } \right],
\end{eqnarray}
so that a straightforward manipulation yields
\begin{eqnarray}
&& \frac{1}{2}\varepsilon _0 {\bf{\hat E}}^{\left( {in} \right)}  \cdot {\bf{\hat E}}^{\left( {in} \right)}  + \frac{1}{{2\mu _0 }}{\bf{\hat B}}^{\left( {in} \right)}  \cdot {\bf{\hat B}}^{\left( {in} \right)}  = \nonumber \\ 
&&   = \frac{{\hbar c^2 }}{{4\left( {2\pi } \right)^3 }}\int {d^3 {\bf{k}}} \int {d^3 {\bf{k}}'} \sum\limits_\nu  {\sum\limits_{\nu '} {\frac{1}{{\sqrt {kk'} }} \cdot } } \nonumber  \\ 
&&   \cdot \left[ {{\bf{e}}_{{\bf{n}}\nu }  \cdot {\bf{e}}_{{\bf{n}}'\nu '}  + \left( {{\bf{n}} \times {\bf{e}}_{{\bf{n}}\nu } } \right) \cdot \left( {{\bf{n}}' \times {\bf{e}}_{{\bf{n}}'\nu '} } \right)} \right] \cdot \nonumber  \\ 
&&   \cdot \left[ {e^{i\left( {{\bf{k}} + {\bf{k}}'} \right) \cdot {\bf{r}}} e^{ - i\left( {\omega  + \omega '} \right)t} \hat g_{\omega {\bf{n}}\nu } \hat g_{\omega '{\bf{n}}'\nu '}  + } \right. \nonumber \\ 
&& + e^{ - i\left( {{\bf{k}} + {\bf{k}}'} \right) \cdot {\bf{r}}} e^{i\left( {\omega  + \omega '} \right)t} \hat g_{\omega {\bf{n}}\nu }^\dag  \hat g_{\omega '{\bf{n}}'\nu '}^\dag   +  \nonumber \\ 
&&   + e^{ - i\left( {{\bf{k}} - {\bf{k}}'} \right) \cdot {\bf{r}}} e^{i\left( {\omega  - \omega '} \right)t} \hat g_{\omega {\bf{n}}\nu }^\dag  \hat g_{\omega '{\bf{n}}'\nu '}  + \nonumber  \\ 
&&  \left. { + e^{i\left( {{\bf{k}} - {\bf{k}}'} \right) \cdot {\bf{r}}} e^{ - i\left( {\omega  - \omega '} \right)t} \hat g_{\omega {\bf{n}}\nu } \hat g_{\omega '{\bf{n}}'\nu '}^\dag  } \right].
\end{eqnarray}
Inserting this expression into Eq.(\ref{F4}) and performing the spatial integration with the help of the delta function we get
\begin{eqnarray}
&& \hat H^{(S)}  = \frac{{\hbar c^2 }}{4}\int {d^3 {\bf{k}}} \sum\limits_\nu  {\sum\limits_{\nu '} } \frac{1}{k} \cdot \nonumber \\ 
&& \cdot \left\{ {\left[ {{\bf{e}}_{{\bf{n}}\nu }  \cdot {\bf{e}}_{ - {\bf{n}}\nu '}  - \left( {{\bf{n}} \times {\bf{e}}_{{\bf{n}}\nu } } \right) \cdot \left( {{\bf{n}} \times {\bf{e}}_{ - {\bf{n}}\nu '} } \right)} \right]} \right. \cdot  \nonumber \\ 
&&   \cdot \left[ {e^{ - i2\omega t} \hat g_{\omega {\bf{n}}\nu } \hat g_{\omega  - {\bf{n}}\nu '}  + e^{i2\omega t} \hat g_{\omega {\bf{n}}\nu }^\dag  \hat g_{\omega  - {\bf{n}}\nu '}^\dag  } \right] +  \nonumber \\ 
&& + \left[ {{\bf{e}}_{{\bf{n}}\nu }  \cdot {\bf{e}}_{{\bf{n}}\nu '}  + \left( {{\bf{n}} \times {\bf{e}}_{{\bf{n}}\nu } } \right) \cdot \left( {{\bf{n}} \times {\bf{e}}_{{\bf{n}}\nu '} } \right)} \right] \cdot  \nonumber \\ 
&& \cdot \left. {\left[ {\hat g_{\omega {\bf{n}}\nu }^\dag  \hat g_{\omega {\bf{n}}\nu '}  + \hat g_{\omega {\bf{n}}\nu } \hat g_{\omega {\bf{n}}\nu '}^\dag  } \right]} \right\} 
\end{eqnarray}
which is easily seen to reduce to 
\begin{equation}
\hat H^{(S)}  = \frac{{\hbar c^2 }}{2}\int {d^3 {\bf{k}}} \frac{1}{k}\sum\limits_\nu  {\left( {\hat g_{\omega {\bf{n}}\nu }^\dag  \hat g_{\omega {\bf{n}}\nu }  + \hat g_{\omega {\bf{n}}\nu } \hat g_{\omega {\bf{n}}\nu }^\dag  } \right)} 
\end{equation}
by using the vector identity
\begin{equation}
\left( {{\bf{A}} \times {\bf{B}}} \right) \cdot \left( {{\bf{A}} \times {\bf{C}}} \right) = \left( {{\bf{A}} \cdot {\bf{A}}} \right)\left( {{\bf{B}} \cdot {\bf{C}}} \right) - \left( {{\bf{A}} \cdot {\bf{B}}} \right)\left( {{\bf{A}} \cdot {\bf{C}}} \right)
\end{equation}
together with the polarization unit vector orthogonality ${\bf{e}}_{{\bf{n}}\nu }  \cdot {\bf{e}}_{{\bf{n}}\nu '}  = \delta _{\nu \nu '}$. Restoring the polar coordinates for the $\bf k$ integration, using the boson commutation relation in the second of Eqs.(\ref{boson}) and neglecting the (divergent) zero-point energy we eventually get
\begin{equation}
\hat H^{(S)}  = \int\limits_0^\infty  {d\omega } \hbar \omega \int {do_{\bf{n}} } \sum\limits_\nu  {\hat g_{\omega {\bf{n}}\nu }^\dag  \hat g_{\omega {\bf{n}}\nu } } .
\end{equation}


\begin{thebibliography}{10}              
\bibitem{Wester} N. Westerberg and R. Bennett,
                 Perturbative light–matter interactions; from first principles to inverse design,
                 Phys. Rep. {\bf 1026}, 1–63 (2023).
\bibitem{Philb0} T. G. Philbin,               
                 Quantum dynamics of the damped harmonic oscillator, 
                 New J. Phys. {\bf 14}, 083043 (2012).               
\bibitem{Huttn1} B. Huttner and S. M. Barnett,             
                 Dispersion and Loss in a Hopfield Dielectric,
                 Europhys. Lett. {\bf 18}, 487-492 (1992).                 
\bibitem{Huttn2} B. Huttner and S. M. Barnett,
                 Quantization of the electromagnetic field in dielectrics,
                 Phys. Rev. A {\bf 46}, 4306-4322 (1992).
\bibitem{Hopfie} J. J. Hopfield,
                 Theory of the Contribution of Excitons to the Complex Dielectric Constant of Crystals,
                 Phys. Rev. {\bf 112}, 1555-1567 (1958).
\bibitem{Fanooo} U. Fano,
                 Atomic Theory of Electromagnetic Interactions in Dense Materials,
                 Phys. Rev. {\bf 103}, 1202-1218 (1956).
\bibitem{Sutto1} L. G. Suttorp and M. Wubs,
                 Field quantization in inhomogeneous absorptive dielectrics,
                 Phys. Rev. A {\bf 70}, 013816 (2004)
\bibitem{Sutto2} L. G. Suttorp and A. J. van Wonderen,
                 Fano diagonalization of a polariton model for an inhomogeneous absorptive dielectric,
                 Europhys. Lett. {\bf 67}, 766–772 (2004).
\bibitem{Kheir1} F. Kheirandish and M. Soltani,
                 Extension of the Huttner-Barnett model to a magnetodielectric medium,
                 Phys. Rev. A {\bf 78}, 012102 (2008).
\bibitem{Matloo} R. Matloob,
                 Electromagnetic field quantization in a linear isotropic dielectric,
                 Phys. Rev. A {\bf 69}, 052110 (2004).
\bibitem{Kheir2} F. Kheirandish and M. Amooshahi,
                 Electromagnetic field quantization in a linear polarizable and magnetizable medium,
                 Phys. Rev. A {\bf 74}, 042102 (2006).
\bibitem{Bhattt} N. A. R. Bhat and J. E. Sipe,
                 Hamiltonian treatment of the electromagnetic field in dispersive and absorptive structured media,
                 Phys. Rev. A {\bf 73}, 063808 (2006).
\bibitem{Amoos1} M. Amooshahi and F. Kheirandish,
                 Electromagnetic field quantization in a magnetodielectric medium with external charges,
                 Phys. Rev. A {\bf 76}, 062103 (2007).              
\bibitem{Sutto3} L. G. Suttorp,                
                 Field quantization in inhomogeneous anisotropic dielectrics with spatio-temporal dispersion,
                 J. Phys. A: Math. Theor. {\bf 40}, 3697–3719 (2007).             
\bibitem{Amoos2} M. Amooshahia,
                 Canonical quantization of electromagnetic field in an anisotropic polarizable and magnetizable medium,
                 J. Math. Phys. {\bf 50}, 062301 (2009).
\bibitem{Grune1} T. Gruner and D. G. Welsch,
                 Green-function approach to the radiation-field quantization for homogeneous and inhomogeneous Kramers-Kronig dielectrics,
                 Phys. Rev. A {\bf 53}, 1818-1829 (1996).
\bibitem{Schee1} S. Scheel, L. Kn{\"o}ll, and D. G. Welsch
                 QED commutation relations for inhomogeneous Kramers-Kronig dielectrics,
                 Phys. Rev. A {\bf 58}, 700-706 (1998).
\bibitem{Dungg1} H. T. Dung, L. Kn{\"o}ll, and D. G. Welsch, 
                 Three-dimensional quantization of the electromagnetic field in dispersive and absorbing inhomogeneous dielectrics,
                 Phys. Rev. A {\bf 57}, 3931-3942 (1998). 
\bibitem{Schee2} S. Scheel and S. Y. Buhmann,
                 Macroscopic quantum electrodynamics - concepts and applications,
                 Acta Phys. Slovaca {\bf 58}, 675-809 (2008)
\bibitem{Buhma1} S. Y. Buhmann, D. T. Butcher and S. Scheel,
                 Macroscopic quantum electrodynamics in nonlocal and nonreciprocal media,
                 New J. Phys. {\bf 14},  083034 (2012).
\bibitem{CheTai} C. T. Tai,
                 Dyadic Green Functions in Electromagnetic Theory
                 IEEE Press (1994).
\bibitem{Chewww} W. C. Chew, 
                 Waves and Fields in Inhomogenous Media, 
                 IEEE Press  (1995).
\bibitem{Buhma2} S. Y. Buhmann and D. G. Welsch,
                 Casimir-Polder forces on excited atoms in the strong atom-field coupling regime,
                 Phys. Rev. A {\bf 77}, 012110 (2008).
\bibitem{Philb1} T. G. Philbin,               
                 Casimir effect from macroscopic quantum electrodynamics,
                 New J. Phys. {\bf 13}, 063026 (2011).            
\bibitem{Buhma3} S. Y. Buhmann,  
                 Dispersion Forces I: Macroscopic Quantum Electrodynamics and Ground-State Casimir, Casimir Polder and van der Waals Forces,
                 Springer-Verlag, Berlin, Heidelberg (2012). 
\bibitem{Buhma4} S. Y. Buhmann, 
                 Dispersion Forces II: Many-Body Effects, Excited Atoms, Finite Temperature and Quantum Friction,
                 Springer-Verlag, Berlin, Heidelberg (2012).         
\bibitem{Butch1} D. T. Butcher, S. Y. Buhmann and S. Scheel1,                 
                 Casimir–Polder forces between chiral objects,
                 New J. Phys. {\bf 14}, 113013 (2012).
\bibitem{Fuchss} S. Fuchs, F. Lindel, R. V. Krems, G. W. Hanson, M. Antezza and S. Y. Buhmann,    
                 Casimir-Lifshitz force for nonreciprocal media and applications to photonic topological insulators,
                 Phys. Rev. A {\bf 96}, 062505 (2017).
\bibitem{Fiedle} J. Fiedler, F. Spallek, P. Thiyam, C. Persson, M. Bostr{\"o}m, M. Walter and S. Y. Buhmann,
                 Dispersion forces in inhomogeneous planarly layered media: A one-dimensional model for effective polarizabilities,
                 Phys. Rev. A {\bf 99}, 062512 (2019).
\bibitem{Schee3} S. Scheel, L. Kn{\"o}ll, and D. G. Welsch and S. M. Barnett,
                 Quantum local-field corrections and spontaneous decay,
                 Phys. Rev. A {\bf 69}, 1590-1597 (1999).                            
\bibitem{Dungg2} H. T. Dung, L. Kn{\"o}ll, and D. G. Welsch, 
                 Spontaneous decay in the presence of dispersing and absorbing bodies: General theory and application to a spherical cavity,
                 Phys. Rev. A {\bf 62}, 053804 (2000).   
\bibitem{Dzsot1} D. Dzsotjan, A. S. Sørensen and M. Fleischhauer,
                 Quantum emitters coupled to surface plasmons of a nanowire: A Green’s function approach,
                 Phys. Rev. B {\bf 82}, 075427 (2010).
\bibitem{Alpegg} F. Alpeggiani, S. D’Agostino and L. C. Andreani,
                 Surface plasmons and strong light-matter coupling in metallic nanoshells,
                 Phys. Rev. B {\bf 86}, 035421 (2012).
\bibitem{Rivera} N. Rivera, I. Kaminer, B. Zhen, J. D. Joannopoulos and M. Soljačić,
                 Shrinking light to allow forbidden transitions on the atomic scale,
                 Science {\bf 353}, 263-269 (2016).
\bibitem{Hemmer} J. L. Hemmerich, R. Bennett, and S. Y. Buhmann,    
                 The influence of retardation and dielectric environments on interatomic Coulombic decay,
                 Nat. Commun. {\bf 9}, 2934 (2018).
\bibitem{Wangg1} S. Wang, G. D. Scholes and L. Y. Hsu,               
                 Quantum dynamics of a molecular emitter strongly coupled with surface plasmon polaritons: A macroscopic quantum electrodynamics approach,
                 J. Chem. Phys. {\bf 151}, 014105 (2019).
\bibitem{Wangg2} S. Wang, M. W. Lee, Y. T. Chuang, G. D. Scholes L. Y. Hsu, 
                 Theory of molecular emission power spectra. I. Macroscopic quantum electrodynamics formalism,
                 J. Chem. Phys. {\bf 153}, 184102 (2020).    
\bibitem{Kosikk} M. Kosik, O. Burlayenko, C. Rockstuhl, I. Fernandez-Corbaton and K, Słowik,       
                 Interaction of atomic systems with quantum vacuum beyond electric dipole approximation,
                 Sci. Rep. {\bf 10}, 5879 (2020)
\bibitem{Khanb1} M. Khanbekyan, L. Kn{\"o}ll, D. G. Welsch,  A. A. Semenov and W. Vogel,
                 QED of lossy cavities: Operator and quantum-state input-output relations,
                 Phys. Rev. A {\bf 72}, 053813 (2005).
\bibitem{Khanb2} M. Khanbekyan, D. G. Welsch, C. Di Fidio and W. Vogel,
                 Cavity-assisted spontaneous emission as a single-photon source: Pulse shape and efficiency of one-photon Fock-state preparation,
                 Phys. Rev. A {\bf 78}, 013822 (2008). 
\bibitem{Dzsot2} D. Dzsotjan, B. Rousseaux, H. R. Jauslin, G. Colas des Francs, C. Couteau and S. Guerin,
                 Mode-selective quantization and multimodal effective models for spherically layered systems,
                 Phys. Rev. A {\bf 94}, 023818 (2016).
\bibitem{Maroci} C. A. Marocico and J. Knoester,
                 Intermolecular resonance energy transfer in the presence of a dielectric cylinder,
                 Phys. Rev. A {\bf 79}, 053816 (2009).
\bibitem{Hakam1} J. Hakami, L. Wang and M. S. Zubairy,             
                 Spectral properties of a strongly coupled quantum-dot–metal-nanoparticle system,
                 Phys. Rev. A {\bf 89}, 053835 (2014).                       
\bibitem{Hakam2} W. Zhang, J. Ren and X. Zhang,       
                 Tunable superradiance and quantum phase gate based on graphene wrapped nanowire,                
                 Opt. Express {\bf 23}, 22347-22361 (2015).   
\bibitem{Kurman} Y. Kurman and I. Kaminer,
                 Tunable bandgap renormalization by nonlocal ultra-strong coupling in nanophotonics,
                 Nat. Phys. {\bf 16 }, 868–874 (2020).
\bibitem{Feistt} J. Feist, A. I. Fernández-Domínguez and F. J. García-Vidal,
                 Macroscopic QED for quantum nanophotonics: emitter-centered modes as a minimal basis for multiemitter problems,
                 Nanophotonics {\bf 10}, 477–489 (2021).
\bibitem{DiGiu1} V. Di Giulio, O. Kfir, C. Ropers and F. Javier García de Abajo,
                 Modulation of Cathodoluminescence Emission by Interference with External Light,
                 ACS Nano 2021 {\bf 15}, 7290-7304 (2021).
\bibitem{Hayun1} A. Ben Hayun, O. Reinhardt, J. Nemirovsky, A. Karnieli, N. Rivera and I. Kaminer,
                 Shaping quantum photonic states using free electrons,
                 Sci. Adv. {\bf 7}, eabe4270 (2021).
\bibitem{Kfirr1} O. Kfir, V. Di Giulio, F. Javier García de Abajo and C. Ropers,
                 Optical coherence transfer mediated by free electrons,
                 Sci. Adv. {\bf 7}, eabf6380 (2021).
\bibitem{Ciatt1} A. Ciattoni, 
                 Fast Electrons Interacting with Chiral Matter: Mirror-Symmetry Breaking of Quantum Decoherence and Lateral Momentum Transfer, 
                 Phys. Rev. Appl. {\bf 19}, 034086 (2023).
\bibitem{Ciatt2} A. Ciattoni, 
                 Quantum interaction of subrelativistic aloof electrons with mesoscopic samples,
                 Phys. Rev. B {\bf 107}, 125403 (2023).
\bibitem{Philbi} T. G. Philbin,               
                 Canonical quantization of macroscopic electromagnetism,
                 New J. Phys. {\bf 12}, 123008 (2010).                   
\bibitem{DiStef} O. Di Stefano, S. Savasta and R. Girlanda,
                 Mode expansion and photon operators in dispersive and absorbing dielectrics,
                 J. Mod. Opt. {\bf 48}, 67-84 (2001)
\bibitem{Dreze1} A. Drezet,
                 Quantizing polaritons in inhomogeneous dissipative systems,
                 Phys. Rev. A {\bf 95}, 023831 (2017).  
\bibitem{Dreze2} A. Drezet,
                 Equivalence between the Hamiltonian and Langevin noise descriptions of plasmon polaritons in a dispersive and lossy inhomogeneous medium,
                 Phys. Rev. A {\bf 96}, 033849 (2017).  
\bibitem{Franke} S. Franke, J. Ren, S. Hughes and M. Richter,
                 Fluctuation-dissipation theorem and fundamental photon commutation relations in lossy nanostructures using quasinormal modes,
                 Phys. Rev. Res. {\bf 2}, 033332 (2020).
\bibitem{Dorie1} V. Dorier, J. Lampart, S. Guérin, and H. R. Jauslin,               
                 Canonical quantization for quantum plasmonics with finite nanostructures,
                 Phys. Rev. A {\bf 100}, 042111 (2019).
\bibitem{Dorie2} V. Dorier, S. Guérin and H. R. Jauslin,
                 Critical review of quantum plasmonic models for finite-size media,
                 Nanophotonics {\bf 9} 3899–3907 (2020).         
\bibitem{Fores1} C. Forestiere and G. Miano,
                 Operative approach to quantum electrodynamics in dispersive dielectric objects based on a polarization-mode expansion,
                 Phys. Rev. A {\bf 106}, 033701 (2022).             
\bibitem{Fores2} C. Forestiere and G. Miano,
                 Integral formulation of the macroscopic quantum electrodynamics in dispersive dielectric objects,
                 Phys. Rev. A {\bf 107}, 063705 (2023).
\bibitem{Naaaa1} D. Y. Na, J. Zhu and W. C. Chew,
                 Diagonalization of the Hamiltonian for finite-sized dispersive media: Canonical quantization with numerical mode decomposition,
                 Phys. Rev. A {\bf 103}, 063707 (2021).
\bibitem{Naaaa2} D. Y. Na, T. E. Roth , J. Zhu,  W. C. Chew and C. J. Ryu,            
                 Numerical framework for modeling quantum electromagnetic systems involving finite-sized lossy dielectric objects in free space,
                 Phys. Rev. A {\bf 107}, 063702 (2023).
\bibitem{Hanson} G. W. Hanson, F. Lindel, S. Y. Buhmann,
                 The Langevin Noise Approach for Lossy Media and the Lossless Limit,
                 J. Opt. Soc. Am. B {\bf 38}, 758-768 (2021).
\bibitem{Kanwal} R. P. Kanwal,
                 Generalized Functions: Theory and Applications 3rd ed.,
                 Birkhäuser, Boston (2004). 
\bibitem{Bornnn} M. Born and E. Wolf,
                 Principle of Optics,
                 Cambridge University Press (2019).
\end{thebibliography}
\end{document}